\documentclass[prd,eqsecnum,twocolumn,amsfonts,amssymb]{revtex4}

\usepackage{graphicx}
\usepackage{bm}
\usepackage{rotating}

\newcommand{\beq}{\begin{equation}}
\newcommand{\eeq}{\end{equation}}
\newcommand{\beqs}{\begin{eqnarray}}
\newcommand{\eeqs}{\end{eqnarray}}
\newcommand{\lsim}{\mathrel{\raisebox{-
.6ex}{$\stackrel{\textstyle<}{\sim}$}}}

\begin{document}

\title{Ultraviolet to Infrared Evolution and Nonperturbative Behavior of 
${\rm SU}(N) \otimes {\rm SU}(N-4) \otimes {\rm U}(1)$ Chiral Gauge Theories}  

\author{Thomas A. Ryttov$^a$ and Robert Shrock$^b$} 

\affiliation{(a) \ CP$^3$-Origins, University of Southern Denmark,
Campusvej 55, Odense, Denmark}

\affiliation{(b) \ C. N. Yang Institute for Theoretical Physics and
Department of Physics and Astronomy, \\
Stony Brook University, Stony Brook, NY 11794, USA }

\begin{abstract}

  We analyze the ultraviolet to infrared evolution and nonperturbative
  properties of asymptotically free ${\rm SU}(N) \otimes {\rm SU}(N-4) \otimes
  {\rm U}(1)$ chiral gauge theories with $N_f$ copies of chiral fermions
  transforming according to $([2]_N,1)_{N-4} + ([\bar 1]_N,[\bar
  1]_{N-4})_{-(N-2)} + (1,(2)_{N-4})_N$, where $[k]_N$ and $(k)_N$ denote the
  antisymmetric and symmetric rank-$k$ tensor representations of SU($N$) and
  the rightmost subscript is the U(1) charge. We give a detailed discussion for
  the lowest nondegenerate case, $N=6$. These theories can exhibit both
  self-breaking of a strongly coupled gauge symmetry and induced dynamical
  breaking of a weakly coupled gauge interaction symmetry due to fermion
  condensates produced by a strongly coupled gauge interaction.  A connection
  with the dynamical breaking of ${\rm SU}(2)_L \otimes {\rm U}(1)_Y$
  electroweak gauge symmetry by the quark condensates $\langle \bar q q\rangle$
  due to color SU(3)$_c$ interactions is discussed. We also remark on
  direct-product chiral gauge theories with fermions in higher-rank tensor
  representations.

\end{abstract}

\maketitle


\section{Introduction}
\label{intro_section}

A problem of basic field-theoretic interest concerns the behavior of strongly
coupled chiral gauge theories. In general, there are two types of chiral gauge
theories, namely those based on a single gauge group and those with a
direct-product ($dp$) gauge group of the form
\beq
G_{dp} = \bigotimes_{i=1}^{N_G} G_i 
\label{gdp}
\eeq
with $N_G \ge 2$.  Strongly coupled direct-product chiral gauge theories are of
particular interest because they can exhibit a phenomenon that cannot occur in
a chiral gauge theory with a single gauge group, namely the induced dynamical
breaking of a weakly coupled gauge symmetry by a different, strongly coupled,
gauge interaction. This phenomenon is important not only from the point of view
of abstract quantum field theory, but also because it actually occurs in
nature. In the Standard Model (SM), with the gauge group $G_{SM}={\rm SU}(3)_c
\otimes G_{EW}$, where the electroweak gauge group is $G_{EW} = {\rm SU}(2)_L
\otimes {\rm U}(1)_Y$, the bilinear quark condensates $\langle \bar q q
\rangle$ produced by the strongly coupled SU(3)$_c$ color gauge interaction
dynamically break $G_{EW}$ to the elctromagnetic gauge symmetry,
U(1)$_{em}$. This breaking contributes terms of the form $g^2 f_\pi^2/4$ and
$(g^2+g'^2)f_\pi^2/4$ to the squared masses of the $W$ and $Z$ bosons, $m_W^2$
and $m_Z^2$, respectively, where $g$ and $g'$ are the SU(2)$_L$ and U(1)$_Y$
gauge couplings, and $f_\pi = 93$ MeV is the pion decay constant.
Thus, although textbook discussions usually mention only the
vacuum expectation value (VEV)
\beq
\langle \phi \rangle_0 = {0 \choose \frac{v}{\sqrt{2}}}
\label{phivev}
\eeq
of the Higgs field $\phi = {\phi^+ \choose \phi^0}$ as the source of
electroweak symmetry breaking in the SM, this breaking really arises from two
different sources, one of which is the Higgs VEV (\ref{phivev}), yielding
$m_W^2 = g^2 v^2/4$ and $m_Z^2 = (g^2+g'^2)v^2/4$, where $v=246$ GeV, and the
other of which is the above-mentioned dynamical contribution due to the
formation of bilinear quark condensates in quantum chromodynamics (QCD).

Although this dynamical breaking of electroweak gauge symmetry by the SU(3)$_c$
color gauge interaction is very small compared with the contribution due to the
VEV of the Higgs field, it is important as a physical example of how, in a
direct-product chiral gauge theory, one strongly coupled gauge interaction can
induce the breaking of a weakly coupled one \cite{weinberg76,tc}. Indeed, a
{\it gedanken} modification of the Standard Model in which the Higgs field is
removed is a perfectly well-defined theory in which the $W$ and $Z$ masses
are entirely due to the dynamical breaking of the electroweak gauge
symmetry by the SU(3)$_c$ interaction \cite{tc,smr}. In the Standard Model, the
SU(3)$_c$ gauge interaction is vectorial, while $G_{EW}$ is chiral, but this
mechanism can also break a vectorial gauge symmetry; in Ref. \cite{smr} it was
shown that in this {\it gedanken} modification of the SM without any Higgs
field, if one reversed the order of the coupling strengths of the non-Abelian
gauge interactions so that the SU(2)$_L$ coupling were much stronger than the
SU(3)$_c$ coupling, then the SU(2)$_L$ gauge interaction would produce bilinear
fermion condensates of quarks and leptons that would break the vectorial
SU(3)$_c$, as well as U(1)$_Y$ and U(1)$_{em}$, \cite{smr}, while preserving
SU(2)$_L$.

Since dynamical symmetry breaking of a weakly coupled gauge symmetry occurs in
nature, as shown by the breaking of electroweak gauge symmetry
$G_{EW}$ by the $\langle \bar q q \rangle$ quark condensates produced by
SU(3)$_c$ gauge interaction, there is a motivation to investigate chiral gauge
theories that can exhibit this phenomenon of the dynamical breaking of a weakly
coupled gauge symmetry by a different, strongly coupled gauge interaction. As
noted above, this requires that one consider theories with direct-product
chiral gauge symmetries.  Some previous studies of strongly coupled chiral
gauge theories with direct-product gauge groups (and without any fundamental
scalar fields) include \cite{smr}-\cite{dpg}, \cite{refs,cgt}.

In this paper we shall analyze chiral gauge theories with the direct-product
gauge group 
\beq
G = {\rm SU}(N) \otimes {\rm SU}(N-4) \otimes {\rm U}(1) \ . 
\label{ggen}
\eeq
This group is of the form (\ref{gdp}) with $N_G=3$, $G_1 = {\rm SU}(N)$, $G_2 =
{\rm SU}(N-4)$, and $G_3 = {\rm U}(1)$. The group (\ref{ggen}) has order
$o(G)$ and rank $rk(G)$ given by 
\beq
o(G)=2N^2-8N+15, \quad rk(G)=2N-5 \ .  
\label{ork}
\eeq
The fermion content of the theory 
consists of $N_f$ copies (``flavors'') of chiral fermions transforming as 
\beq
([2]_N,1)_{N-4} + ([\bar 1]_N, [\bar 1]_{N-4})_{-(N-2)} + 
(1,(2)_{N-4})_N \ , 
\label{fermions}
\eeq
where the meaning of the notation
\beq
(R_1,R_2)_q
\label{rrq}
\eeq
is as follows: the first and second entries refer to the representation $R_1$
of $G_1={\rm SU}(N)$ and $R_2$ of $G_2={\rm SU}(N-4)$, and the subscript $q$ is
the U(1) charge of the given fermion. The symbols $[k]_N$ and $(k)_N$ denote
the $k$-fold antisymmetric and symmetric tensor representations of SU($N$),
respectively, and $R_i=1$ denotes a singlet of $G_i$, where $i=1$ or $i=2$.
The fermion fields are denoted explicitly as 
\beqs
&& ([2]_6,1)_2: \ \psi^{ij}_{p,L} \ , \cr\cr
&& ([\bar 1]_6: \ [\bar 1]_2)_{-4}: \ \chi_{i,\alpha,p,L} \ , \cr\cr
&& (1,(2)_2)_6: \ \omega^{\alpha\beta}_{p,L} \ , 
\label{explicitfermions_n6}
\eeqs
where $i,j$ are SU($N$) group indices, $\alpha,\beta$ are SU($N-4$) group
indices, and $p$ is a copy (flavor) index, running from 1 to $N_f$. We
exclude the trivial value $N_f=0$, because it does not produce a chiral gauge
theory, but instead just a set of three decoupled pure gauge theories.  There
are no bare fermion masses in the theory, since they are forbidden by the
chiral gauge symmetry.  Without loss of generality, we write the fermions as
left-handed.  This theory is free of anomalies in gauged currents, as is
necessary for renormalizability, and is also free of global anomalies and mixed
gauge-gravitational anomalies \cite{barr2015,anomcal}. 

We note two equivalent theories with the same gauge group, (\ref{ggen}).  The
first of these has all of the representations of the
left-handed chiral fermions in (\ref{fermions}) conjugated.  The second has the
representations of SU($M$) conjugated relative to those of SU($N$), i.e., its
fermion content consists of $N_f$ copies of the set 
\beq
([2]_N,1)_{N-4} + ([\bar 1]_N, [1]_{N-4})_{-(N-2)} + 
(1,(\bar 2)_{N-4})_N \ , 
\label{fermions_sunm4_reversed}
\eeq
Since these theories are equivalent to 
(\ref{ggen}) with (\ref{fermions}), it will suffice to study only the latter.

This model is of particular interest for the following reason.  A natural
construction of a chiral gauge theory with a non-Abelian gauge group uses
(left-handed chiral) fermions transforming according to an antisymmetric or
symmetric rank-$k$ tensor representation of the gauge group, together with the
requisite number of fermions transforming according to the conjugate
fundamental representation, so as to yield zero gauge anomaly.  The simplest of
these uses $k=2$, so let us focus on these theories with $k=2$. With a special
unitary gauge group, there are two such constructions: (i) $G={\rm SU}(N)$ and
chiral fermion content consisting of $N_f$ copies of the set 
\beq
[2]_N + (N-4) \, [\bar 1]_N 
\label{a2fb}
\eeq
and (ii) $G={\rm SU}(M)$ and chiral fermion content consisting of $N_f$ 
copies of the set
\beq
(2)_M + (M+4) \, [\bar 1]_M \ . 
\label{s2fb}
\eeq
A basic question in the analysis of chiral gauge theories is whether one can
combine these two separate single-gauge-group theories (i) and (ii) into a
single chiral gauge theory with a direct-product gauge group that contains
${\rm SU}(N) \otimes {\rm SU}(M)$ such that it is again anomaly-free.  The
answer is yes, if we set $M=N-4$, and the theory
(\ref{ggen}) with (\ref{fermions}) provides an explicit realization of this
combination. Indeed, not only does this theory successfully combine the two
separate chiral gauge symmetries SU($N$) and SU($M$) in an anomaly-free manner;
it also incorporates a third gauge symmetry, namely the U(1). 

A general classification of chiral gauge theories with direct-product gauge
groups was given in Ref. \cite{dpg}.  In this classification, a factor group
$G_i$ is labelled as $G_c$ if it has complex representations and $G_r$ if it
has (only) real or pseudoreal representations. If a group $G_c$ has no gauge
anomaly from any of its representations, then it was denoted as $G_{cs}$, where
the subscript $s$ stands for ``safe''. In this classification, if $N \ge 7$,
then the gauge group (\ref{ggen}) is of the form $(G_c,G_c,G_c)$. In contrast,
if $N=6$, then the second factor group is SU(2), which has (pseudo)real
representations, so that the $N=6$ special case of (\ref{ggen}) is of the form
$(G_c,G_r,G_c)$ in this classification.

In accordance with the order of labelling of the $G_i$ factor groups, we denote
the corresponding running gauge couplings as $g_1(\mu)$ for $G_1 = {\rm
  SU}(N)$, $g_2(\mu)$ for $G_2 = {\rm SU}(N-4)$, and $g_3(\mu)$ for $G_3 = {\rm
  U}(1)$, where $\mu$ is the Eucidean energy/momentum reference scale where
$g_i(\mu)$ is measured.  We further define $\alpha_i(\mu) = g_i(\mu)^2/(4\pi)$
and $a_i(\mu) = g_i(\mu)^2/(16\pi^2)$, with $i=1,2,3$.  (The argument $\mu$
will sometimes be suppressed in the notation.)  As usual with a U(1) gauge
interaction, the U(1) charge assignments in (\ref{fermions}) involve an
implicit normalization convention; the physics is unchanged if one redefines 
$q_f \to \lambda q_f$ for each fermion $f$ and $g_3 \to \lambda^{-1}g_3$, 
since only the product $q_f g_3$ appears in the U(1) covariant derivative. 

Each of the two non-Abelian gauge interactions is required to be asymptotically
free (AF), because this enables us to calculate the corresponding beta
functions self-consistently at a high scale $\mu=\mu_{_{UV}}$ in the deep
ultraviolet (UV) region, where they are weakly coupled. These beta functions
then describe the running of the non-Abelian couplings toward the infrared (IR)
at small $\mu$, where these couplings become larger.  Since we are interested
in the nonperturbative behavior of the non-Abelian gauge interactions, we will
assume the U(1) gauge interaction to be weakly coupled at the initial reference
scale $\mu_{_{UV}}$; owing to the property that the beta function for this U(1)
interaction is non-asymptotically free, the U(1) coupling $\alpha_3(\mu)$
becomes even weaker as $\mu$ decreases below $\mu_{_{UV}}$ and hence can be
treated perturbatively in the full range $\mu < \mu_{_{UV}}$ under
consideration here.

In addition to the phenomenon of a strongly coupled gauge interaction inducing
the dynamical breaking of a different gauge symmetry, a chiral gauge theory can
also exhibit a different phenomenon in which a strongly coupled gauge
interaction corresponding to a given gauge symmetry produces fermion
condensates that break this gauge symmetry itself
\cite{weinberg76,tumbling}.  In particular, for a given gauge
interaction corresponding to the non-Abelian gauge group $G_i$, as $\mu$
decreases from $\mu_{_{UV}}$ and $\alpha_i(\mu)$ grows, it may become large
enough at a certain scale, which we will denote as $\mu=\Lambda_1$, to produce
a fermion condensate that breaks the gauge symmetry $G_i$ to a subgroup $H_i
\subset G_i$. The fermions involved in this condensate gain dynamical masses of
order $\Lambda_1$ and are integrated out of the low-energy effective
field theory (EFT) that describes the physics as $\mu$ decreases below
$\Lambda_1$.  The gauge bosons in the coset space $G_i/H_i$ pick up dynamical
masses of order $g_i(\Lambda_1) \, \Lambda_1$ and are also integrated out of
the low-energy effective theory. This low-energy theory has a gauge coupling
inherited from the UV theory, but since the fermion and gauge boson content is
different, this gauge coupling runs according to a different beta function.
Then this process of self-breaking of a gauge can repeat at one or more lower
scales. The final low-energy effective field theory may be a
vectorial theory that confines and produces fermion condensates with associated
spontaneous chiral symmetry breaking (S$\chi$SB) but no further gauge
self-breaking.

Besides being of abstract field-theoretic interest, this mechanism of gauge
self-breaking has been used in constructions and studies of reasonably
ultraviolet-complete models of dynamical electroweak symmetry breaking (EWSB)
and fermion mass generation \cite{at94}-\cite{sml}, \cite{gen}. In these
constructions, one starts with an asymptotically free chiral gauge theory that
undergoes either self-breaking or a combination of self-breaking and induced
symmetry breaking in a sequence of three different scales, $\Lambda_1 >
\Lambda_2 > \Lambda_3$, with an associated breaking of the UV chiral gauge
symmetry $G_{UV} \to H_1 \to H_2 \to H_3$, where the $H_3$ symmetry is
vectorial. At a lower scale $\Lambda_T$ of order 1 TeV, the $H_3$ gauge
interaction confines and produces condensates that break $G_{EW}$.  It also
produces a spectrum of $H_3$-singlet bound states. Gauge bosons in the coset
space $G_{UV}/H_1$ gain dynamical masses of order $\Lambda_1$, while gauge
bosons in the coset spaces $H_1/H_2$ and $H_2/H_3$ gain dynamical masses of
order $\Lambda_2$ and $\Lambda_3$, respectively. Exchanges of these three
different types of massive vector bosons produce the three generations of quark
and lepton masses.  More complicated exchanges can also produce light neutrino
masses via an appropriate seesaw mechanism \cite{nt}. This scenario has the
potential to naturally explain the generational hierarchy in fermion masses,
which reflects the hierarchy of self-breaking scales $\Lambda_i$,
$i=1,2,3$. This construction is also an ultraviolet completion of low-energy
effective Lagrangians for dynamical EWSB that use four-fermion operators
\cite{etc} and predicts the coefficients of these four-fermion operators. 

Our theory does not include any fundamental scalar fields. Thus, the pattern of
possible dynamical gauge symmetry breaking depends only on the gauge and
fermion content, and the initial values of the gauge couplings at the reference
scale $\mu_{_{UV}}$.  This is in contrast with theories in which gauge symmetry
breaking is produced by VEVs of Higgs fields, because in these latter theories,
the nature of the symmetry breaking depends on various parameters in the Higgs
potential, which can be chosen at will, subject to the constraint that this
Higgs potential should be bounded from below \cite{isb,barr2018}.

An alternate application of strongly coupled chiral gauge theories was to
efforts at modelling the quarks and leptons as composites of more fundamental
fermions, commonly called preons. This involved a scenario in which it was
envisioned that the strongly coupled gauge interaction would produce
confinement of the preons in gauge-singlet composite fermions, but no
spontaneous chiral symmetry breaking. The presumed absence of S$\chi$SB was
necessary in order for the composite fermions to be very light compared to the
inverse of the spatial compositeness scale $\Lambda_{comp.} = 1/r_{comp.}$. For
this purpose, theories were constructed that satisfied certain matching
conditions of chiral symmetries between preons and the composite fermions
\cite{thooft1979,preons}. In the present paper we will focus on studying
possible patterns of bilinear fermion condensate formation and resultant
dynamical gauge symmetry breaking in the strongly coupled gauge theory
(\ref{ggen}) with (\ref{fermions}) and (\ref{gn6})-(\ref{fermions_n6}) rather
than on possible scenarios with light composite fermions.

In addition to our analysis of the general theory (\ref{ggen}) with
(\ref{fermions}), we will study the $N=6$ special case in detail. 
This $N=6$ theory, with the gauge group
\beq
G_{N=6} = {\rm SU}(6) \otimes {\rm SU}(2) \otimes {\rm U}(1) \ , 
\label{gn6}
\eeq
is of particular interest because it is the lowest non-degenerate member of
this family. (If $N=5$, then the SU($N-4$) group is trivial).  It is also
special in two related aspects, namely that (i) as mentioned above, the
resultant second factor group is SU(2), with (pseudo)real representations, in
contrast to the situation for $N \ge 7$, where the SU($N-4$) group has complex
representations; and (ii) the symmetric rank-2 tensor representation $(2)_2$ of
SU(2) is the adjoint representation.  The fermion content for this $N=6$
theory, comprised of $N_f$ copies of
\beq
([2]_6,1)_2+([\bar 1]_6, [\bar 1]_2)_{-4}+(1,(2)_2)_6 \ , 
\label{fermions_n6}
\eeq
can also be conveniently expressed in terms of the dimensionalities of
the representations as
\beq
 (15,1)_2 + (\bar 6,2)_{-4} + (1,3)_6 \ . 
\label{fermions_n6_dim}
\eeq
Owing to property (ii) above, we will often use the equivalent isovector 
notation ${\vec \omega}_{p,L}$ for the $\omega^{\alpha\beta}_{p,L}$ fermion.

Thus, the theory (\ref{ggen}) with (\ref{fermions}) and, in particular, the
$N=6$ special case, provide valuable theoretical laboratories for the study 
of nonperturbative properties of chiral gauge theories, including 
self-breaking of a strongly coupled chiral gauge symmetry, induced breaking of
a weakly coupled gauge symmetry by a strongly coupled gauge interaction, and
the sequential construction of low-energy effective field theories. 
This paper is organized as follows.  The general methods used in our analysis
are described in Section \ref{methods_section}. In Section \ref{ggen_section}
we analyze the UV to IR evolution, possible fermion condensation channels, and
corresponding gauge symmetry breaking patterns of the theory (\ref{ggen}) with
(\ref{fermions}).  In Sections \ref{n6_section}-\ref{n6_both_section} we
present a detailed analysis of the $N=6$ theory.  Some remarks on related
constructions of direct-product chiral gauge theories with fermions in
higher-rank tensor representations are given in Section
\ref{related_constructions_section}. Our conclusions are contained in Section
\ref{conclusions_section}.


\section{Renormalization-Group Evolution and Fermion Condensates}
\label{methods_section}


\subsection{Beta Functions}

In this section we discuss the general methods that are used for our analysis.
We first explain our application of the renormalization group (RG).  Recall our
labelling conventions given above for the gauge couplings, namely $g_1(\mu)$
for SU($N$), $g_2(\mu)$ for SU($N-4$), and $g_3(\mu)$ for U(1).  The evolution
of the three gauge couplings $g_i(\mu)$, or equivalently, the corresponding
$\alpha_i(\mu)$ with $i=1,2,3$, is determined by the RG beta functions
\beq
\beta_{G_i} = \frac{d\alpha_i(\mu)}{d\ln\mu} \ .
\label{beta_gi}
\eeq
These have the series expansions  
\beqs
\beta_{G_i} & = &-8\pi a_i \bigg [ b^{(G_i)}_{1\ell,i}a_i +
\sum_{j=1}^3 b^{(G_i)}_{2\ell;ij} a_i a_j \cr\cr
& + & \sum_{j,k=1}^3 b^{(G_i)}_{3\ell;ijk}a_ia_ja_k + ... \bigg ] \ ,
\label{beta}
\eeqs
where an overall minus sign is extracted, the dots $...$ indicate
higher-loop terms, and there is no sum on repeated $i$ indices in the square
bracket. Here, $b^{(G_i)}_{1\ell,i}$ is the one-loop ($1\ell$)
coefficient, multiplying $a_i$ inside the square bracket in (\ref{beta});
$b^{(G_i)}_{2\ell;ij}$ is the two-loop coefficient, multiplying $a_ia_j$ in the
square bracket, and so forth for higher-loop terms.  The one-loop
coefficients $b^{(G_i)}_{1\ell,i}$ are scheme-independent.

We focus on the beta functions for the two non-Abelian gauge interactions,
since these determine the upper bound on $N_f$ and are relevant for the
formation of various possible fermion condensates as $\alpha_1(\mu)$ or
$\alpha_2(\mu)$ become large in the infrared.  The one-loop coefficients in
Eq. (\ref{beta}) are
\beq
b^{({\rm SU}(N))}_{1\ell,1} = \frac{1}{3}\Big [ 11N - 2N_f(N-3) \Big ] \ , 
\label{b1_sun}
\eeq
\beq
b^{({\rm SU}(N-4))}_{1\ell,2} = \frac{1}{3}\Big [ 11(N-4) - 2N_f(N-1) \Big ] \
, 
\label{b1_sunm4}
\eeq
and 
\beq
b^{({\rm U}(1))}_{1\ell,3} = -\frac{4}{3} \, N_f \, N(N-1)(N-3)(N-4) \ .
\label{b1_u1}
\eeq
As mentioned before, we assume that the U(1) gauge interaction is weakly
coupled at the UV reference scale $\mu_{_{UV}}$; then its gauge coupling
decreases as $\mu$ decreases from the UV to the IR, and hence can be treated
perturbatively.

The requirements that the SU($N$) and SU($N-4$) gauge interactions must be
asymptotically free are that $b^{({\rm SU}(N))}_{1\ell,1} > 0$ and
$b^{({\rm SU}(N-4))}_{1\ell,2} > 0$. These impose the respective upper limits
$N_f < N_{f,b1z}$ and $N_f < N_{f,b1z}'$, where
\beq
N_{f,b1z} = \frac{11N}{2(N-3)}
\label{nfb1z_sun}
\eeq
and
\beq
N_{f,b1z}' = \frac{11(N-4)}{2(N-1)} \ , 
\label{nfb1z_sunm4}
\eeq
where we use a prime to indicate the upper limit on $N_f$ from the condition
$b^{({\rm SU}(N-4))}_{1\ell} > 0$.
The upper bound (\ref{nfb1z_sunm4}), is more restrictive than the upper bound
(\ref{nfb1z_sun}), as is clear, since the difference
\beq
N_{f,b1z} - N_{f,b1z}' = \frac{33(N-2)}{(N-1)(N-3)}
\label{nfb1z_dif}
\eeq
is positive for all of the relevant values of $N$ under consideration
here. Hence, we restrict
\beq
N_f < \frac{11(N-4)}{2(N-1)} \ .
\label{nf_upper}
\eeq
The (nonzero) values of $N_f$ that are allowed by the inequality 
(\ref{nf_upper}) depend on $N$ and are as follows:
\begin{enumerate}

\item $1 \le N_f \le 2$ if $6 \le N \le 7$ 
\item $1 \le N_f \le 3$ if $8 \le N \le 12$
\item $1 \le N_f \le 4$ if $13 \le N \le 34$
\item $1 \le N_f \le 5$ if $N_f \ge 35$.

\end{enumerate}
As $N \to \infty$, the upper limit on $N_f$ (formally generalized to a
non-negative real number) approaches 11/2, thus allowing physical integral
values up to 5, inclusive, as indicated above.

In general, the set of equations (\ref{beta}) is comprised of three coupled
nonlinear first-order ordinary differential equations for the quantities
$\alpha_i$, $i=1,2,3$.  The solutions for the three $\alpha_i(\mu)$ depend on
$N_f$ and the three initial values $\alpha_i(\mu_{_{UV}})$ at the UV reference
scale $\mu_{_{UV}}$.  Since we do not assume that the group (\ref{ggen}) is
embedded in a single gauge group higher in the UV, we may choose these initial
values $\alpha_i(\mu_{_{UV}})$ arbitrarily, subject to the constraint that for
$\mu=\mu_{_{UV}}$, the values are sufficiently small that the perturbative
calculation of the beta functions $\beta_{\alpha_i}$ are self-consistent.
To leading order, i.e., to one-loop order, the differential equations making up
this set decouple from each other, and one has the simple solution for each
$i=1,2,3$:
\beq
\alpha_i(\mu_1)^{-1} = \alpha_i(\mu_2)^{-1}-\frac{b^{(G_i)}_{1\ell,i}}{2\pi} \,
\ln\Big ( \frac{\mu_2}{\mu_1} \Big ) \ ,
\label{alfsol}
\eeq
where we take $\mu_1 < \mu_2$. 

At the level of two loops and higher, due to the fact that each of the fermions
has nonzero U(1) charge and one of the fermions, $\chi_{i,\alpha,p,L}$, is a
nonsinglet under both of the non-Abelian gauge groups, there are mixed terms
$a_ia_j$, $a_ia_ja_k$, etc., that involve different gauge interactions, in the
three beta functions $\beta_{\alpha_i}$, so that the three beta functions
become coupled differential equations.  In view of the mixing terms in
(\ref{beta}) at the two-loop level, it is natural to focus first
on two special cases, namely those in which one of the non-Abelian gauge
interactions is much stronger than the other.  This can be arranged by
specifying appropriate initial values of $\alpha_1(\mu_{_{UV}})$ and
$\alpha_2(\mu_{_{UV}})$ at the UV scale $\mu_{_{UV}}$. In these two cases, one
can neglect the two-loop term that mixes these two non-Abelian gauge
interactions in Eq. (\ref{beta}), so that, to two-loop level, these
interactions decouple, and the corresponding beta functions have the form, to
this level,
\beq
\beta_{\alpha_1}= \frac{d\alpha_1}{d\ln\mu} =
-8\pi a_1 \, \Big [b^{({\rm SU}(N))}_{1\ell,1}a_1 + 
b^{({\rm SU}(N))}_{2\ell;11} a_1^2 \Big ]
\label{beta_sun_2loop}
\eeq
and
\beq
\beta_{\alpha_2}= \frac{d\alpha_2}{d\ln\mu} = 
-8\pi a_2 \, \Big [b^{({\rm SU}(N-4))}_{1\ell,2}a_2 + 
b^{({\rm SU}(N-4))}_{2\ell;22} a_2^2 \Big ] \ ,
\label{beta_su2_2loop}
\eeq
where the one-loop coefficients $b^{({\rm SU}(N))}_{1\ell,1}$ and 
$b^{({\rm SU}(N-4))}_{1\ell,2}$ were given above in 
Eqs. (\ref{b1_sun})-(\ref{b1_u1}), and the two-loop coefficients are
\beqs
&& b^{({\rm SU}(N))}_{2\ell;11} = \frac{1}{6N} 
\bigg [ 68N^3-N_f(N-3)(29N^2-3N-12) \bigg ]  \cr\cr
&&
\label{b2_sun}
\eeqs
and
\begin{widetext}
\beq
b^{({\rm SU}(N-4))}_{2\ell;22} = \frac{1}{6(N-4)} 
\bigg [ 68(N-4)^3 -N_f(N-1)(29N^2-229N+440) \bigg ] \ .
\label{b2_sunm4}
\eeq
\end{widetext}

Both of these two-loop coefficients for the non-Abelian gauge couplings are
positive for small $N_f$ and decrease with increasing $N_f$, eventually passing
through zero to negative values.  We denote the values of $N$ (formally
generalized from positive integers $N \ge 6$ to positive real numbers) at which
$b^{({\rm SU}(N))}_{2\ell;11}$ and $b^{({\rm SU}(N-4))}_{2\ell;22}$ pass
through zero as $N_{f,b2z}^{({\rm SU}(N))}$ and $N_{f,b2z}^{({\rm
    SU}(N-4))}$. These are
\beq
N_{f,b2z}^{({\rm SU}(N))} = \frac{68N^3}{(N-3)(29N^2-3N-12)}
\label{nfb2z_sun}
\eeq
and
\beq
N_{f,b2z}^{({\rm SU}(N-4))} = \frac{68(N-4)^3}{(N-1)(29N^2-229N+440)} \ . 
\label{nfb2z_sunm4}
\eeq
As $N \to \infty$, $N_{f,b2z}^{({\rm SU}(N))}$ approaches $68/29 = 2.34483$ 
from above, while $N_{f,b2z}^{({\rm SU}(N-4))}$ approaches the same value from
below. 

With these inputs, we can investigate the presence or absence of an IR zero in
the respective two-loop beta functions for the SU($N$) and SU($N-4$) theories.
The two-loop beta function for SU($N$)
has no IR zero for $N_f=1$ or $N_f=2$; it does have an IR zero for higher
values of $N_f$, as allowed by the asymptotic freedom requirement for a fixed
$N$. With a given $N$, for the range of $N_f$ such that
$b^{({\rm SU}(N))}_{1\ell,1} > 0$ and $b^{({\rm SU}(N))}_{2\ell;11} < 0$, the
IR zero of the two-loop SU($N$) beta function occurs at
\beqs
\alpha_{1,IR,2\ell} &=& \frac{8\pi N[11N-2N_f(N-3)]}
{N_f(N-3)(29N^2-3N-12)-68N^3} \ . \cr\cr
&&
\label{alfir_2loop_sun}
\eeqs
Similarly, given a value of $N$, for the range of $N_f$ such that $b^{({\rm
    SU}(N-4))}_{1\ell,2} > 0$, while $b^{({\rm SU}(N-4))}_{2\ell;22} < 0$, the
two-loop SU($N-4$) beta function has an IR zero at
\beqs
\alpha_{2,IR,2\ell} &=& \frac{8\pi(N-4)[11(N-4)-2N_f(N-1)]}
{N_f(N-1)(29N^2-229N+440)-68(N-4)^3} \ . \cr\cr
&& 
\label{alfir_2loop_sunm4}
\eeqs
As $N \to \infty$, the rescaled IRFP values of the SU($N$) and SU($N-4$) gauge
interactions have the same limit:
\beqs
\lim_{N \to \infty} \alpha_{1,IR,2\ell}N &=& 
\lim_{N \to \infty} \alpha_{2,IR,2\ell}N \cr\cr\
&=& \frac{8\pi(11-2N_f)}{29N_f-68} \ . 
\label{alfir_2loop_sun_Ninfinity}
\eeqs
We will analyze the UV to IR evolution using these beta functions below. 


\subsection{Global Flavor Symmetries}

The theory (\ref{ggen}) with (\ref{fermions}) has the classical global
flavor (cgb) symmetry 
\beq
G_{cgb} = {\rm U}(N_f)_{\psi} \otimes{\rm U}(N_f)_{\chi} 
\otimes{\rm U}(N_f)_{\omega} \ , 
\label{gglobal_classical}
\eeq
where, for each fermion $f=\psi^{ij}_{p,L}$, $\chi_{i,\alpha,p,L}$, and ${\vec
  \omega}_{p,L}$, the elements of the group ${\rm U}(N_f)_f$ act on the flavor
indices $p$, leaving all gauge indices unchanged. Each ${\rm U}(N_f)_f$ factor
group in (\ref{gglobal_classical}) can equivalently be written as ${\rm
  SU}(N_f)_f \otimes {\rm U}(1)_f$. The instantons present in the SU($N$) gauge
sector break both of the global abelian symmetries U(1)$_\psi$ and 
U(1)$_\chi$.  Separately, the instantons in the SU($N-4$) gauge sector break
both the U(1)$_\chi$ and U(1)$_{\omega}$ symmetries.  

There are two special cases that will be of particular interest, namely the
respective cases in which one non-Abelian gauge interaction is much stronger
than the other.  First, let us consider the case in which the SU($N$) gauge
interaction is much stronger than the SU($N-4$) gauge interaction, which, like
the U(1) interaction, is weakly coupled.  In this theory, the effects of
instantons in the SU($N-4$) gauge sector are exponentially suppressed and can
be neglected \cite{b_minus_ell}.  
Although the SU($N$) instantons break the global U(1)$_\psi$ and
U(1)$_\chi$ flavor symmetries, one can construct a current which is a linear
combination of the U(1)$_\psi$ and U(1)$\chi$ currents and is conserved in
the presence of the SU($N$) instantons (see, e.g., 
Section V of \cite{cgtakfb}), which we denote as U(1)$_{\psi\chi}$.  
The effective non-anomalous global flavor (gb) symmetry of this theory 
is thus $G_{gb}={\rm SU}(N_f)_\psi \otimes {\rm SU}(N_f)_\chi \otimes
{\rm U}(1)_{\psi\chi} \otimes {\rm U}(N_f)_\omega$. 
Similarly, in the other case, in which the SU($N$) and U(1) gauge interactions
are weak, and the SU($N-4$) gauge interaction is strong, the effects of SU($N$)
instantons are exponentially suppressed and are negligible.  Although the
SU($N-4$) instantons break the global U(1)$_\omega$ and U(1)$_\chi$ flavor
symmetries, one can construct a current which is a linear combination of the
U(1)$_\omega$ and U(1)$\chi$ currents and is conserved in the presence of the
SU($N$) instantons, which we denote as U(1)$_{\omega\chi}$.  The effective
non-anomalous global flavor symmetry of this theory is thus
$G_{gb} = {\rm SU}(N_f)_\omega \otimes {\rm SU}(N_f)_\chi \times
{\rm U}(1)_{\omega\chi} \otimes {\rm U}(N_f)_\psi$. 


\subsection{UV to IR Evolution and Fermion Condensates} 

We next discuss the UV to IR evolution of this theory and the general analysis
of possible fermion condensate formation in various channels. We begin with the
two respective cases in which one of the two non-Abelian gauge interactions is
much stronger than the other and then remark on the case where both are present
with comparable strength. Let us denote the dominant coupling as
$\alpha_i(\mu)$.

As the reference scale $\mu$ decreases below $\mu_{_{UV}}$, the coupling
$\alpha_i(\mu)$ for this interaction increases.  There are two general
possibilities for the associated beta function, $\beta_{\alpha_i}$: (i) it does
not have an IR zero or (ii) it has an IR zero. In the first case, (i), the
coupling continues to increase with decreasing $\mu$ until it eventually
exceeds the range where it can be calculated with the perturbative beta
function. This can then lead to the formation of (bilinear) fermion
condensates.  In the second case, let us denote the value of $\alpha_i$ at this
IR zero as $\alpha_{IR}$, and consider a possible condensation channel,
\beq
R \times R' \to R_c \ ,  
\label{condensation_channel} 
\eeq
where $R$ and $R'$ denote fermion representations under the strongly coupled
gauge symmetry $G_i$, and $R_c$ denotes the representation of the condensate
under $G_i$. Assuming that this is an attractive channel, we denote the
minimal critical coupling for condensation in this channel as $\alpha_{cr}$.
If the beta function does not have an IR zero, then $\alpha_i$ will certainly
exceed $\alpha_{cr}$ as $\mu$ decreases to some scale.  If the beta function
$\beta_{\alpha_i}$ does have an IR zero, then there are two subcases: (iia)
$\alpha_{IR} \ge \alpha_{cr}$ and (iib) $\alpha_{IR} < \alpha_{cr}$. In case
(iia), the condensate can form, similarly to case (i), while in case (iib),
this condensate will not form.  For the possible condensation channel
(\ref{condensation_channel}), an approximate measure of its attractiveness
(motivated by iterated one-gluon exchange) is
\beq
\Delta C_2 = C(R) + C(R') - C(R_c) \ , 
\label{deltac2}
\eeq
where $C_2(R)$ is the quadratic Casimir invariant for the representation $R$
\cite{casimir}.  Among several possible fermion condensation channels, the one
with the largest (positive) value of $\Delta C_2$ is commonly termed the most
attractive channel (MAC) and is the one that is expected to occur. 

Approximate solutions of Schwinger-Dyson equations for the fermion propagator
in a vectorial theory have shown that if one starts with a massless fermion, it
follows that if $\alpha > \alpha_{cr}$, where $3 \alpha_{cr} C_2(R)/\pi=1$,
then the Schwinger-Dyson equation has a solution with a dynamically generated
mass, indicating spontaneous chiral symmetry breaking and associated bilinear
fermion condensate formation \cite{alm}. In a vectorial gauge theory such as
quantum chromodynamics, the condensate is a gauge-singlet, so $\Delta C_2 = 2
C_2(R)$. Hence, one can write the condition for the critical coupling in the
form that can be taken over for a chiral gauge theory, namely $3 \alpha_{cr}
\Delta C_2/(2\pi)=1$, so that
\beq
\alpha_{cr} = \frac{2\pi}{3\Delta C_2 } \ . 
\label{alfcrit}
\eeq
Because this is based on a rough approximation (an iterated one-gluon exchange
approximation to the Schwinger-Dyson equation), it is used only as a
rough estimate.

Since without loss of generality we write all fermions as left-handed, the
Lorentz-invariant bilinears involving two fermion fields $f_L$ and $f'_L$ are 
of the form $f^T_L C f'_L$, where $C$ is the Dirac charge-conjugation matrix
satisfying $C\gamma_\mu C^{-1} = -(\gamma_\mu)^T$.
If $f_L$ and $f'_L$ transform according to the same representation $R_1$ of a
symmetry group $G_1$ and $R_2$ of a symmetry group $G_2$, 
then we may write the bilinear fermion operator product abstractly as 
\beq
f^T_{{\cal R},p,L} C f_{{\cal R},p',L} \ , 
\label{ff}
\eeq
where gauge group indices are suppressed in the notation, ${\cal R}$ denotes
the representations under the gauge groups, and, as before, $p$
and $p'$ are flavor indices.  From the property $C^T = -C$ together with the
anticommutativity of fermion fields, it follows that the bilinear fermion
operator product (\ref{ff}) is symmetric under interchange of the order of
fermion fields and therefore is symmetric in the overall product
\beq
 \prod_{i} (R_i \times R_i)] \, R_{fl} \ , 
\label{symprod}
\eeq
where $R_{fl}$ abstractly denotes the symmetry property under interchange of
flavors \cite{dpg}. For our theory, with its two non-Abelian
groups, this means that the fermion bilinears are of the form 
\beq
(s,s,s), \quad (s,a,a), \quad (a,s,a), \ \ {\rm or} \ \ (a,a,s) \ , 
\label{sss}
\eeq
where here $s$ and $a$ indicate symmetric and antisymmetric, and the three
entries refer to the representations $R_1$ of $G_1$, $R_2$ of $G_2$, and
$R_{fl}$.

If, as $\mu$ decreases through a scale $\Lambda_1$ and the coupling
$\alpha_i(\mu)$ of the strongly coupled gauge interaction corresponding to the
factor group $G_i$ increases beyond $\alpha_{cr}$ for the condensation channel
(\ref{condensation_channel}) and the condensate forms, then the fermions
involved in the condensate gain dynamical masses of order $\Lambda_1$ and are
integrated out of the low-energy effective field theory that describes the
physics as $\mu$ decreases below $\Lambda_1$.  If this condensate either
self-breaks the $G_i$ symmetry or produces induced breaking of a weakly coupled
gauge symmetry $G_j$ to a respective subgroup $H_i \subset G_i$ or $H_j \subset
G_j$, then the gauge bosons in the respective coset spaces $G_i/H_i$ or
$G_j/H_j$ pick up dynamical masses of order $g_i(\Lambda_1) \Lambda_1$ or
$g_j(\Lambda_1) \, \Lambda_1$, respectively.  Hence, like the fermions with
dynamically generated masses, these now massive vector bosons are integrated
out of the low-energy effective field theory applicable as $\mu$ decreases
below $\Lambda$.


\section{ Theory with General $N$} 
\label{ggen_section}

In this section we analyze possible fermion condensation channels in the
general-$N$ theory (\ref{ggen}) with fermion content (\ref{fermions}). 


\subsection{${\rm SU}(N)$ Gauge Interaction Dominant } 
\label{sun_dominant_subsection}

We begin by focusing on the case where the SU($N$) gauge interaction is much
stronger than the SU($N-4$) (and U(1)) gauge interactions.  This theory is
labelled SUND, standing for ``SU($N$)-dominant''.  Although we keep $\alpha_2$
and $\alpha_3$ nonzero, we note parenthetically that if one were to set
$\alpha_2=\alpha_3=0$, then the resultant theory would be the $k=2$ special
case of a family of chiral gauge theories analyzed in Ref. \cite{cgtakfb} with
a single gauge group $G={\rm SU}(N)$ and an anomaly-free content of chiral
fermions transforming as $[k]_N$ and $n_{\bar F}$ copies of $[\bar 1]_N$, where
$n_{\bar F} = (N-3)!(N-2k)/[(N-k-1)!(k-1)!]$ (plus SU($N$)-singlet
fermions). Since the SU($N$) gauge interaction is asymptotically free,
$\alpha_1(\mu)$ increases as $\mu$ decreases from the initial reference scale
$\mu_{_{UV}}$ in the UV.  We focus on the subset of values of $N_f$ such that
the beta function $\beta_{\alpha_1}$ either has no IR zero at the two-loop
level or has an IRFP at a sufficiently large value that fermion condensation
can take place.

There are three possible (bilinear) fermion condensation channels. We give
shorthand names to these based on the fermions involved in each condensate. 
The first is the $\psi\chi$ channel, 
\begin{widetext}
\beq
\psi\chi: \quad ([2]_N,1)_{N-4} \times ([\bar 1]_N,[\bar 1]_{N-4})_{-(N-2)} 
\to ([1]_N,[\bar 1]_{N-4})_{-2} \ , 
\label{a2_fbar_to_f_sun}
\eeq
\end{widetext}
with associated condensate
\beq
\langle \psi^{ij \ T}_{p,L} C \chi_{j, \beta,p',L} \rangle \ , 
\label{psi_chi_condensate_sun}
\eeq
where $i,j$ are SU($N$) group indices and $\beta$ is an SU($N-4$) group index. 
The condensate
(\ref{psi_chi_condensate_sun}) transforms as the fundamental ($[1]_N$)
representation of SU($N$) and the conjugate fundamental ($[\bar 1]_{N-4}$) 
representation of SU($N-4$), so it self-breaks SU($N$) to SU($N-1$) and
produces an induced breaking of the weakly coupled SU($N-4$) to SU($N-5$).
Since the condensate
(\ref{psi_chi_condensate_sun}) has a nonzero U(1) charge, $q_{\psi\chi}=-2$, it
also breaks U(1). Thus, here the residual gauge symmetry in the effective field
theory that is applicable as $\mu$ decreases below $\Lambda_1$ is 
\beq
{\rm SU}(N-1) \otimes {\rm SU}(N-5) \ . 
\label{hinv}
\eeq
If $N=6$, then the residual gauge symmetry is just SU(5). 
For this channel we calculate 
\beq
(\Delta C_2)_{\psi \chi} = C_2([2]_N) = \frac{(N-2)(N+1)}{N} \ . 
\label{Delta_C2_a2_fbar_to_f_sun}
\eeq
For this and other possible fermion condensation channels, we record these
properties in Table \ref{condensate_properties_sun}. This table refers to the
possible initial condensation patterns at the highest condensation scale;
subsequent evolution further into the infrared is discussed below.

The second possible channel is the $\psi\psi$ channel, 
\begin{widetext}
\beq
\psi\psi: \quad ([2]_N,1)_{N-4} \times ([2]_N,1)_{N-4} \to 
([4]_N,1)_{2(N-4)} \ , 
\label{a2_a2_to_a4_sun}
\eeq
\end{widetext}
Note that $[4]_N \approx [\overline{N-4}]_N$, where 
$R \approx R'$ means that the representations $R$ and $R'$ are
equivalent.  The associated condensate is 
\beq
\epsilon_{...k\ell mn} \langle \psi^{k \ell \ T}_{p,L} C 
\psi^{m n}_{p',L} \rangle \ , 
\label{psi_psi_condensate_sun}
\eeq
where the antisymmetric tensor $\epsilon_{...k\ell mn}$ has $N$ indices, four
of which are indicated explicitly, with the rest implicit. From the general
group-theoretic analysis in \cite{kim82,cgtakfb}, it follows that since the 
condensate (\ref{psi_psi_condensate_sun}) transforms as a $[4]_N$ of SU($N$),
it breaks SU($N$) to ${\rm SU}(N-4) \otimes {\rm
  SU}(4)$. Since the $\psi^{k\ell}_{p,L}$ are singlets under the original
SU($N-4$) group in (\ref{ggen}), this condensate is obviously invariant under
this SU($N-4$). Furthermore, since this condensate has a nonzero U(1) charge
(namely, $q_{\psi\psi}=2(N-4)$), it breaks the U(1) gauge symmetry. Hence, the
condensate (\ref{psi_psi_condensate_sun}) breaks $G$ to
\beq
[{\rm SU}(N-4) \otimes {\rm SU}(4)] \otimes {\rm SU}(N-4) \ . 
\label{psi_psi_invariance_group_ggen}
\eeq
where we have inserted brackets to distinguish the two different SU($N-4$)
groups. The measure of attractiveness for this condensation channel is
\beq
(\Delta C_2)_{\psi\psi} = 2C_2([2]_N) - C_2([4]_N) = \frac{4(N+1)}{N} \ . 
\label{Delta_C2_a2_a2_to_a4_sun}
\eeq

The third possible channel is the $\chi\chi$ channel, 
\begin{widetext}
\beq
\chi\chi: \quad ([\bar 1]_N, [\bar 1]_{N-4})_{-(N-2)} \times 
                ([\bar 1]_N, [\bar 1]_{N-4})_{-(N-2)} \to 
([\bar 2]_N,[\bar 2]_{N-4})_{-2(N-2)} \ , 
\label{chi_chi_channel}
\eeq
\end{widetext}
with associated condensate
\beq
\epsilon^{...mn} \epsilon^{...\alpha\beta} 
\langle \chi_{m \alpha,p,L}^T C \chi_{n,\beta,p',L} \rangle \ , 
\label{chi_chi_condensate_sun}
\eeq
where $\epsilon^{...mn}$ and $\epsilon^{...\alpha\beta}$ are antisymmetric
tensors under SU($N$) and SU($N-4$), respectively, with two indices shown
explicitly and the rest understood implicitly.  From the general 
group-theoretic analysis \cite{kim82,cgtakfb}, it follows that since the
condensate (\ref{chi_chi_condensate_sun}) transforms as a 
$[\bar 2]_N$ representation of SU($N$), it breaks SU($N$) to
${\rm SU}(N-2) \otimes {\rm SU}(2)'$, and similarly, since it transforms as a 
$[\bar 2]_{N-4}$ representation of SU($N-4$), it breaks SU($N-4$) to 
${\rm SU}(N-6) \otimes {\rm SU}(2)''$.  Here we append a single prime to the
first SU(2) and a double prime to the second SU(2) to distinguish them and
also to disguish them from the SU(2) group of the $N=6$ theory (\ref{gn6}). 
Since the condensate (\ref{chi_chi_condensate_sun})
carries a nonzero U(1) charge
(namely, $q_{\chi\chi}=-2(N-2)$), it breaks the U(1) gauge symmetry. Thus, 
this condensate (\ref{chi_chi_condensate_sun}) breaks $G$ to the group
\beq
[{\rm SU}(N-2) \otimes {\rm SU}(2)'] \otimes 
[{\rm SU}(N-6) \otimes {\rm SU}(2)''] \ , 
\label{chi_chi_invariance_group_ggen}
\eeq
where the square brackets here are inserted to indicate the origin of the
different factor groups from the original SU($N$) and SU($N-4$) factor groups
in (\ref{ggen}). We find 
\beq
(\Delta C_2)_{\chi\chi} = 2C_2([\bar 1]_N) - C_2([\bar 2]_N) = \frac{N+1}{N} \
. 
\label{Delta_C2_fbar_fbar_to_a2bar_sun}
\eeq

From these results we calculate the relative attractiveness of these three
possible fermion condensation channels in this SU($N$)-dominant case. We 
compute the differences 
\beq
(\Delta C_2)_{\psi\chi}-(\Delta C_2)_{\psi\psi} = \frac{(N-6)(N+1)}{N}
\label{psichi_minus_psipsi}
\eeq
and
\beq
(\Delta C_2)_{\psi\psi}-(\Delta C_2)_{\chi\chi} = \frac{3(N+1)}{N} \ , 
\label{psipsi_minus_chichi}
\eeq
whence 
\beq
(\Delta C_2)_{\psi\chi}-(\Delta C_2)_{\chi\chi} = \frac{(N-3)(N+1)}{N} \ , 
\label{psichi_minus_chichi}
\eeq
and the ratios
\beq
\frac{(\Delta C_2)_{\psi\chi}}{(\Delta C_2)_{\psi\psi}}= \frac{N-2}{4} 
\label{psichi_over_psipsi}
\eeq
and
\beq
\frac{(\Delta C_2)_{\psi\psi}}{(\Delta C_2)_{\chi\chi}} = 4 \ , 
\label{psipsi_over_chichi}
\eeq
whence
\beq
\frac{(\Delta C_2)_{\psi\chi}}{(\Delta C_2)_{\chi\chi}} = N-2  \ . 
\label{psichi_over_chichi}
\eeq
Therefore, in this SU($N$)-dominant case with $N \ge 7$, the $\psi\chi$ channel
is the MAC, with greater attractiveness than the $\psi\psi$ channel, which, in
turn, is more attractive than the $\chi\chi$ channel. Summarizing:
\begin{widetext} 
\beq
{\rm SU}(N)-{\rm dominant \ with} \ N \ge 7 \ \Longrightarrow \ \psi\chi \  
{\rm channel \ is \ the \ MAC}.
\label{psi_chi_mac_sun}
\eeq
\end{widetext} 
One interesting feature of these comparisons is
that the ratio $(\Delta C_2)_{\psi\psi}/(\Delta C_2)_{\chi\chi}$ is independent
of $N$. As is evident from these results, in the lowest non-degenerate case, 
namely $N=6$, the $\psi\chi$ and $\psi\psi$ channels are equally
attractive, and are a factor 4 more attractive
than the $\chi\chi$ channel. Thus, 
\begin{widetext} 
\beq
{\rm SU}(N)-{\rm dominant \ with} \ N =6 \ \Longrightarrow \ \psi\chi \ 
{\rm and} \ \psi\psi \ {\rm channels \ are \ the \ MACs}.
\label{su6macs}
\eeq
\end{widetext} 

We focus here on the range $N \ge 7$; a detailed analysis of the $N=6$ case
will be given below.  Since the $\psi\chi$ channel is the MAC, it is expected
that as the Euclidean reference scale $\mu$ decreases below a value that we
denote as $\Lambda_1$, the coupling $\alpha_1(\mu)$ increases sufficiently to
cause condensation in this channel. This condensation self-breaks SU($N$) to
SU($N-1$) and breaks the weakly coupled gauge symmetry SU($N-4$) to SU($N-5$)
and also the abelian symmetry U(1).  Without loss of generality, we may choose
the SU($N$) group index $i$ in the condensate (\ref{psi_chi_condensate_sun}) to
be $i=N$ and the SU($N-4$) group index to be $\alpha=N-4$.  The condensate
(\ref{psi_chi_condensate_sun}) also spontaneously breaks the global flavor
group $G_{gb}$ for this theory, producing a set of Nambu-Goldstone bosons
(NGBs).  Earlier works in related chiral gauge theories have studied the
resultant change in counts of the UV versus IR degrees of freedom
\cite{cgtakfb}, \cite{df}-\cite{cgtsab}. Here we focus on the dynamical
self-breaking and induced breaking of gauge symmetries, together with the
construction of resultant low-energy effective field theories. The fermions
involved in the condensate (\ref{psi_chi_condensate_sun}), namely
$\psi^{Nj}_{p,L} = -\psi^{jN}_{p,L}$ with $1 \le j \le N-1$ and
$\chi_{j,N-4,p',L}$ with $1 \le j \le N-1$, gain dynamical masses of order
$\Lambda_1$. The $2N-1$ SU($N$) gauge bosons in the coset space ${\rm
  SU}(N)/{\rm SU}(N-1)$ gain dynamical masses of order $g_1(\Lambda_1)
\Lambda_1$, while the $(2N-9)$ \ SU($N-4$) gauge bosons in the coset space
${\rm SU}(N-4)/{\rm SU}(N-5)$ gain dynamical masses of order $g_2(\Lambda_1)
\Lambda_1$.  Finally, the U(1) gauge boson picks up a dynamical mass of order
$g_3(\Lambda_1) \Lambda_1$. These massive fields are integrated out of the
low-energy effective field theory that describes the physics as the reference
scale $\mu$ decreases below $\Lambda_1$.

This low-energy effective field theory that is applicable as $\mu$ decreases
below $\Lambda_1$ is invariant under the gauge symmetry
(\ref{hinv}). The massless gauge-nonsinglet fermion content of
this EFT consists of

\begin{enumerate}

\item $\psi^{ij}_{p,L}$ 
with $1 \le i,j \le N-1$, $1 \le p \le N_f$, forming $N_f$ copies of a 
$([2]_{N-1},1)$ representation under the group (\ref{hinv}),

\item $\chi_{j,\beta,p',L}$ with $1 \le j \le N-1$, 
$1 \le \beta \le N-5$, and $1 \le p' \le N_f$, comprising $N_f$ copies of the
$([\bar 1]_{N-1},[\bar 1]_{N-5})$ representation of 
(\ref{hinv}), and

\item $\omega^{\alpha\beta}_{p,L}$ with $1 \le \alpha, \beta \le N-5$ and 
$1 \le p \le N_f$, comprising $N_f$ copies of $(1,(2)_{N-5})$.

\end{enumerate}
(We do not list the U(1) charges, since there is no U(1) gauge symmetry in this
low-energy effective theory.)  The condensation process then repeats, with the
$\psi \chi$ condensation channel again being the MAC in this ${\rm SU}(N-1)
\otimes {\rm SU}(N-5)$ theory.  One can treat the successive self-breakings and
induced dynamical breakings iteratively at the various steps.


\subsection{${\rm SU}(N-4)$ Gauge Interaction Dominant, $N \ge 7$} 
\label{sum_dominant_subsection}

Here we analyze the case in which the SU($N-4$) gauge interaction is strongly
coupled and dominates over the SU($N$) gauge interaction (as well as the weakly
coupled U(1) gauge interaction).  We restrict to the range $N \ge 7$ here and
will consider the $N=6$ in detail below. It will sometimes be convenient to use
the quantity $M=N-4$ as defined before.  We will denote this theory as SUMD,
standing for ``SU($M$)-dominant''. If we were to completely turn off the
SU($N$) and U(1) gauge interactions, then this theory would be equivalent to a
chiral gauge theory with an SU($M$) gauge group, and $N_f$ flavors of chiral
fermions transforming according to the anomaly-free set $(2)_M$ + $M+4$ copies
of $[\bar 1]_M$, which has been studied in \cite{df}-\cite{cgtsab}. However,
here we do not completely turn off the SU($N$) or U(1) gauge interactions.

There are two possible (bilinear) fermion condensation channels.  The first is 
\beqs
\chi\omega: \quad 
&& ([\bar 1]_N, [\bar 1]_{N-4})_{-(N-2)} \times 
(1,(2)_{N-4})_N \to \cr\cr
&&\to ([\bar 1]_N,[1]_{N-4})_2 \ , 
\label{chi_omega_sunm4}
\eeqs
with associated condensate
\beq
\langle \chi_{i,\alpha,p,L}^T C \omega^{\alpha\beta}_{p',L} \rangle \ ,
\label{chi_omega_condensate_sunm4}
\eeq
where $i$ is an SU($N$) group index and $\alpha, \ \beta$ are 
${\rm SU}(N-4)$ group
indices.  The value of $\Delta C_2$ for this condensation, as produced by the
SU($N-4$) gauge interaction, is 
\beqs
(\Delta C_2)_{\chi\omega,SUMD} &=& 
C_2((2)_{N-4}) = \frac{(N-2)(N-5)}{N-4} \cr\cr
&& {\rm in} \ {\rm SU}(N-4) \ . 
\label{Delta_C2_chi_omega_sunm4}
\eeqs
This condensate transforms as $([\bar 1]_N,[1]_{N-4})_2$ and hence self-breaks
SU($N-4$) to SU($N-5$) and produces induced breaking of the weakly coupled
symmetries SU($N$) to SU($N-1$) and of U(1).  It leaves invariant the same
residual gauge symmetry, (\ref{hinv}), as the $\psi\chi$ condensate
(\ref{psi_chi_condensate_sun}), which is the MAC for the SU($N$)-dominant case
(\ref{hinv}). By convention,
one may choose the SU($N-4$) index $\beta$ in the condensate
(\ref{chi_omega_condensate_sunm4}) to be $\beta=N-4$ and the SU($N$) index $i$
to be $i=N$.  Then the fermions $\chi_{N,\alpha,p,L}$ and
$\omega^{\alpha,N-4}_{p',L}$ with $1 \le \alpha \le N-4$, $1 \le p,p' \le N_f$
involved in the condensate pick up dynamical masses of order $\Lambda_1$.  The
dynamical mass generation for the SU($N$) and SU($N-4$) gauge bosons in the
respective coset spaces ${\rm SU}(N)/{\rm SU}(N-1)$ and ${\rm SU}(N-4)/{\rm
  SU}(N-5)$ is the same as described above in the SU($N$)-dominant scenario, as
is the dynamical mass generation for the U(1) gauge boson.

A second possible condensation channel is the $\chi\chi$ channel 
(\ref{chi_chi_channel}), with associated condensate 
(\ref{chi_chi_condensate_sun}).  This condensate breaks $G$ to the group given
above in Eq. (\ref{chi_chi_invariance_group_ggen}). 
The measure of attractiveness of this 
condensation channel, as produced by the ${\rm SU}(M)={\rm SU}(N-4)$ 
gauge interaction, is 
\beqs
(\Delta C_2)_{\chi\chi} &=& 2C_2([\bar 1]_M) - C_2([\bar 2]_M) = \frac{M+1}{M} 
\cr\cr
&=& \frac{N-3}{N-4}  \ . 
\label{Delta_C2_chi_chi_sunm4}
\eeqs

Comparing the attractiveness measure of the channels (\ref{chi_omega_sunm4})
and (\ref{chi_chi_channel}), we calculate the difference
\beq
(\Delta C_2)_{\chi\omega} - (\Delta C_2)_{\chi\chi} = \frac{N^2-8N+13}{N-4}
\label{chiomega_minus_chichi_sunm4}
\eeq
and the ratio
\beq
\frac{(\Delta C_2)_{\chi\omega}}{(\Delta C_2)_{\chi\chi}} = 
\frac{(N-2)(N-5)}{N-3} \ . 
\label{chiomega_over_chichi_sunm4}
\eeq
For the range $N \ge 6$, the difference 
$(\Delta C_2)_{\chi\omega} - (\Delta C_2)_{\chi\chi}$ is positive and,
equivalently, the ratio $(\Delta C_2)_{\chi\omega}/
(\Delta C_2)_{\chi\chi} > 1$.  Hence, the $\chi\omega$ channel is always more
attractive than the $\chi\chi$ channel in this SU($N-4$)-dominant case. Thus,
\begin{widetext}
\beq
{\rm SU(N-4)-dominant \ with} \ N \ge 7: \ \Longrightarrow \ \chi\omega \ 
{\rm channel \ is \ the \ MAC}.
\label{chi_omega_mac_sunm4}
\eeq
\end{widetext}
In addition to breaking gauge symmetries, the MAC condensate
(\ref{chi_omega_condensate_sunm4}) spontaneously breaks the global symmetry
$G_{gb}$ for this theory, yielding a set of NGBs. Here we focus on the gauge
symmetry breaking.  We have restricted to the range $N \ge 7$; as will be
discussed below, the MAC is different in the special case $N=6$, where it is
the $\omega\omega$ channel.

Although the $\chi\chi$ channel is not the MAC, we comment on its symmetry
properties.  It breaks SU($N-4$) to ${\rm SU}(N-6) \otimes
{\rm SU}(2)$ and also breaks U(1), since the condensate has nonzero U(1)
charge $q_{\chi\chi}=-2(N-2)$. In terms of SU($N-4$), the associated
condensate has the form
\beq
\epsilon^{...\alpha \beta} 
\langle \chi_{i,\alpha,p,L}^T C \chi_{j,\beta,p',L}\rangle \ , 
\label{chi_chi_condensate_sunm4_initial}
\eeq
where $\epsilon^{...\alpha \beta}$ is an antisymmetric SU($N-4$) tensor and we
have indicated $N-6$ of the indices implicitly with dots.  
For this $\chi\chi$ channel, as regards the SU($N$) and flavor
symmetry, there are two channels and corresponding condensates.  The
(\ref{chi_chi_channel}) 
channel that involves an antisymmetric structure for
SU($N$) group indices is
\beqs
&& ([\bar 1]_N, [\bar 1]_{N-4})_{-(N-2)} \times 
([\bar 1]_N,[\bar 1]_{N-4})_{-(N-2)} \to \cr\cr
&& \to ([\bar 2]_N,[\bar 2]_{N-4})_{-2(N-2)} \ , 
\label{chi_chi_sunm4_sun_asym}
\eeqs
with corresponding condensate 
\beq
\epsilon^{...mn} \epsilon^{...\alpha\beta} 
\langle \chi_{m,\alpha,p,L}^T C \chi_{n,\beta,p',L}\rangle  \ . 
\label{chi_chi_condensate_sunm4_sun_asym}
\eeq
Here $\epsilon^{...mn}$ is an antisymmetric tensor under SU($N$), 
$\epsilon^{...\alpha\beta}$ was defined, and we indicate the rest of
the indices in each tensor implicitly with dots. This condensate is
automatically symmetrized in the flavor indices $p$ and $p'$ and is of the form
$(a,a,s)$ in the classification of Ref. \cite{dpg}.
The (\ref{chi_chi_channel}) channel that involves a symmetric structure for
SU($N$) group indices is
\beqs
&& ([\bar 1]_N, [\bar 1]_{N-4})_{-(N-2)} \times 
([\bar 1]_N,[\bar 1]_{N-4})_{-(N-2)} \to \cr\cr
&& \to ((\bar 2)_N,[\bar 2]_{N-4})_{-2(N-2)} \ , 
\label{chi_chi_sunm4_sun_sym}
\eeqs
with corresponding condensate 
\beqs
\epsilon^{...\alpha\beta} 
\langle \chi_{i,\alpha,p,L}^T C \chi_{j,\beta,p',L}\rangle - (p \leftrightarrow
p') \ . 
\label{chi_chi_condensate_sunm4_sun_sym}
\eeqs
Because this condensate is antisymmetrized in flavor indices, it is 
automatically symmetric in SU($N$) group indices and is thus of the form 
$(s,a,a)$ in the classification of Ref. \cite{dpg}. 


\subsection{SU($N$) and SU($N-4$) Gauge Interactions of Comparable Strength}
\label{both_subsection}

Finally, we analyze the situation in which the SU($N$) and SU($N-4$) gauge
interactions are of comparable strength at the scale relevant for the initial
condensation.  We restrict to $N \ge 7$ here and will discuss the $N=6$ theory
below.  The value of $\Delta C_2$ for the most attractive channel, $\psi\chi$,
in the SU($N$)-dominant case was given in Eq. 
(\ref{Delta_C2_a2_fbar_to_f_sun}), and the value of 
$\Delta C_2$ for the MAC $\chi\omega$ in the SU($N-4$)-dominant case was given
in Eq. (\ref{Delta_C2_chi_omega_sunm4}) above. The difference is
\beqs
&& (\Delta C_2)_{\psi\chi, SUND} - (\Delta C_2)_{\chi\omega, SUMD} = 
\frac{4(N-2)}{N(N-4)} \ . \cr\cr
&& 
\label{dc2_dc2_nge7}
\eeqs
Since this is positive for the relevant range of $N$ considered here, it
follows that, as the reference scale decreases and the SU($N$) and SU($N-4$)
couplings increase, the minimal value of $\alpha$ for condensation is reached
first for the SU($N$)-induced $\psi\chi$ condensate, at a scale $\mu$ that we
may again denote $\Lambda_1$, where $\alpha_1(\Lambda_1)$ exceeds $\alpha_{cr}$
for the $\psi\chi$ condensation.  At a slightly lower scale, $\Lambda_1' \lsim
\Lambda_1$, the SU($N-4$) gauge interaction, of comparable strength, increases
through the slightly larger critical value for condensation in the $\chi\omega$
channel.  These condensates both break the gauge symmetry in the same way, to
the residual subgroup ${\rm SU}(N-1) \otimes {\rm SU}(N-5)$, as given in
Eq. (\ref{hinv}). We have described the fermions and gauge bosons that gain
dynamical masses from the $\psi\chi$ and $\chi\omega$ condensations above, and
we combine these results here By convention, one may choose the SU($N$) index
$i$ and the SU($N-4$) index $\alpha$ in the $\psi\chi$ condensate $\langle
\psi^{ij \ T}_{p,L} C \chi_{j,\alpha,p',L}\rangle$ in Eq.
(\ref{psi_chi_condensate_sun}) to be $i=N$ and $\alpha=N-4$, respectively.  The
fermions involved in this condensate are then $\psi^{Nj}_{p,L}$ and
$\chi_{j,N-4,p',L}$ with $1 \le j \le N-1$.  These gain dynamical masses of
order $\Lambda_1$.  The $2N-1$ SU($N$) gauge bosons in the coset ${\rm
  SU}(N)/{\rm SU}(N-1)$ and the $2M-1=2N-9$ SU($M$) gauge bosons in the coset
${\rm SU}(M)/{\rm SU}(M-1)={\rm SU}(N-4)/{\rm SU}(N-5)$ gain dynamical masses
of order $\simeq g_1(\Lambda_1)\Lambda_1$ and $\simeq g_2(\Lambda_1)\Lambda_1$,
while the U(1) gauge boson gains a dynamical mass $\simeq g_3(\Lambda_1)
\Lambda_1$. A vacuum alignment argument \cite{weinberg76,vacalign} suggests
that the condensation process would be such as to preserve the maximal residual
gauge symmetry, with gauge group of the largest order, thereby minimizing the
number of gauge bosons that pick up masses. In the present case, one can use
this argument to infer that in the condensate $\langle \chi_{i,\alpha,p,L}^T C
\omega^{\alpha\beta}_{p',L}\rangle$ in Eq. (\ref{chi_omega_condensate_sunm4}),
the SU($N$) index is the same as the unmatched index in the $\langle \psi^{ij \
  T}_{p,L} C \chi_{j,\alpha,p',L}\rangle$ condensate, namely $i=N$, and the
$\beta$ index is the same as the unmatched SU($N-4$) index $\alpha$ in the
$\psi\chi$ condensate, namely $N-4$, so that these two condensates break the
initial UV gauge group $G$ in the same way, to the subgroup ${\rm SU}(N-1)
\otimes {\rm SU}(N-5)$ in Eq. (\ref{hinv}). Then the fermions
involved in the $\chi\omega$ condensate, $\chi_{N,\alpha,p,L}$ and
$\omega^{\alpha\beta}_{p',L}$ with $1 \le \alpha \le N-4$ and $\beta=N-4$, gain
dynamical masses of order $\Lambda_1$ and $\Lambda_1'$.

The resultant low-energy effective field theory that describes the physics as
the reference scale $\mu$ decreases below $\Lambda_1'$ contains the following
massless fermions that are nonsinglets under the residual gauge group 
${\rm SU}(N-1) \otimes {\rm SU}(N-5)$:

\begin{enumerate}

\item $\psi^{ij}_{p,L}$ 
with $1 \le i,j \le N-1$, $1 \le p \le N_f$, forming $N_f$ copies of a 
$([2]_{N-1},1)$ representation under the group (\ref{hinv}),

\item 
$\chi_{j,\alpha,p',L}$ with $1 \le j \le N-1$, 
$1 \le \beta \le N-5$, and $1 \le p' \le N_f$, forming $N_f$ copies of the
$([\bar 1]_{N-1},[\bar 1]_{N-5})$ representation of 
(\ref{hinv}), and

\item 
$\omega^{\alpha\beta}_{p',L}$ with $1 \le \alpha, \ \beta \le N-5$, forming
$N_f$ copies of the $(1,(2)_{N-5})$ representation of the group 
(\ref{hinv})

\end{enumerate}
This theory also includes certain massless fermions that are singlets under the
gauge group (\ref{hinv}), e.g., $\chi_{N,N-4,p,L}$. 


\section{$N=6$ Theory}
\label{n6_section}


\subsection{Beta Function and Constraints on $N_f$}
\label{n6_general_subsection}

In this section we study the lowest nondegenerate case of the chiral gauge
theory (\ref{ggen}) with the fermions (\ref{fermions}), namely the $N=6$
theory, for which the fermion content was given in
Eq. (\ref{fermions_n6}). From the general formulas (\ref{b1_sun}) and
(\ref{b1_sunm4}), it follows that the one-loop coefficients for the SU(6) and
SU(2) gauge interactions in this theory are
\beq
b^{({\rm SU}(6))}_{1\ell,1} =2(11-N_f)  
\label{b1_su6}
\eeq
and
\beq
b^{({\rm SU}(2))}_{1\ell,2} = \frac{2}{3}\Big (11-5N_f \Big ) \ . 
\label{b1_su2}
\eeq
Substituting $N=6$ into the upper bound on $N_f$ in Eq. (\ref{nf_upper}), we
find that $N_f < 11/5$, i.e., for physical integral values, 
\beq
N=6 \ \Longrightarrow \ N_f=1, \ 2 \ , 
\label{nf12}
\eeq
in accord with the general result given in Section \ref{methods_section}. 
When discussing the $N_f=1$ case, we will suppress the flavor indices in the
notation, since they are all the same.

For the study of the UV to IR evolution of this theory, we substitute $N=6$
into the general formulas (\ref{b2_sun}) and (\ref{b2_sunm4}) to obtain the 
two-loop coefficients in the SU(6) and SU(2) beta functions, which are
\beq
b^{({\rm SU}(6))}_{2\ell;11} = \frac{1}{2}\Big ( 816-169N_f \Big ) 
\label{b2_su6}
\eeq
and
\beq
b^{({\rm SU}(2))}_{2\ell;22} = \frac{1}{6}\Big ( 272 - 275N_f \Big ) \ . 
\label{b2_su2}
\eeq
%


\section{$N=6$ Theory with SU(2) Gauge Interaction Dominant} 
\label{su2_dominant_subsection}


\subsection{RG Evolution from UV} 

As before, it is natural to begin by analyzing the UV to IR evolution in the
case where one non-Abelian gauge interaction is much stronger than the other.
We start with the situation in which the SU(2) gauge interaction is much
stronger than the SU(6) interaction, so that, to first approximation, we may
treat the SU(6) (as well as U(1)) gauge interaction perturbatively. By analogy
with our notation above, this will be denoted as the SU2D case, where again, D
stands for ``dominant''. Then, since $b^{({\rm SU}(2))}_{1\ell,2} > 0$ while
$b^{({\rm SU}(2))}_{2\ell,22} < 0$, the two-loop beta function
$\beta_{\alpha_2}$ for the SU(2) gauge interaction has an IR zero at
\beqs
\alpha_{2,IR,2\ell} &=& -\frac{4\pi b^{({\rm SU}(2))}_{1\ell,2}}
                                   {b^{({\rm SU}(2))}_{2\ell;22}} \cr\cr
&=& \frac{16\pi(11-5N_f)}{275N_f-272} \ . 
\label{alfir_2loop_su2}
\eeqs
For $N_f=1$, $\alpha_{2,IR,2\ell}=32\pi = 100.5$, while for $N_f=2$,
$\alpha_{2,IR,2\ell}=8\pi/139 = 0.181$. The IRFP value for $N_f=1$ is too
large for the two-loop calculation to be considered to be quantitatively
accurate, but it does indicate that the theory becomes strongly coupled in the
IR.  The IRFP value for $N_f=2$ is considerably smaller than the estimates of
the critical values $\alpha_{cr}$ for any of the three attractive condensation
channels (which will be given below). Hence, this theory with $N_f=2$ is
expected to evolve in the IR limit to an exact IR fixed point (IRFP) in a
scale-invariant and conformally-invariant non-Abelian Coulomb phase (NACP),
without any spontaneous chiral symmetry breaking or associated fermion
condensate formation. We therefore focus on the $N_f=1$ case. Since the flavor 
subscripts $p, p'$ are always equal to 1, they are suppressed in the notation. 


\subsection{Condensation at Scale $\Lambda_1$ } 

We proceed to determine the most attractive channel for the formation of
bilinear condensates of SU(2)-nonsinglet fermions in this $N_f=1$ case. There
are, {\it a priori}, several possible channels. The first is
\beq
\omega\omega: \quad (1,Adj)_6 \times (1,Adj)_6 \to (1,1)_{12} \ , 
\label{3x3_to_1_su2}
\eeq
where $Adj$ is the adjoint (triplet) representation of SU(2) and the notation
follows Eq. (\ref{rrq}). The shorthand name for this channel, $\omega \omega$,
follows from the condensate, which is
\beq
\langle {\vec \omega}^T_{L} C \, \cdot {\vec \omega}_{L}\rangle \ .
\label{omega_omega_condensate_su2}
\eeq
In terms of dimensions of the SU(2) representations, this channel has the form
$3 \times 3 \to 1$.  The measure of attractiveness of this channel due to the
strongly coupled SU(2) gauge interaction is 
\beq
\Delta C_2 = 2C_2(Adj) = 4 \quad {\rm for} \ 3 \times 3 \to 1 \ {\rm in} \ 
{\rm SU}(2) \ . 
\label{Delta_C2_3x3_to_1_su2}
\eeq
This is the most attractive channel:
\begin{widetext} 
\beq
{\rm SU}(2)-{\rm dominant} \ \Longrightarrow \ \omega\omega \ {\rm channel \ is \ 
the \ MAC}.
\label{omega_omega_mac_su2}
\eeq
\end{widetext} 
The rough estimate of the minimal
(critical) coupling $\alpha_2(\mu)=\alpha_{cr}$ for this channel is given by
Eq. (\ref{alfcrit}) as $\alpha_{cr} \simeq \pi/6 = 0.5$. Since the condensate
involves the SU(6)-singlet fermion ${\vec \omega}_L$, it obviously preserves
the SU(6) gauge symmetry.  As a scalar product of the isovector ${\vec
  \omega}_L$ with itself, this condensate is also invariant under the strongly
coupled SU(2) gauge symmetry. Because the condensate has a nonzero U(1) charge
(namely, $q=12$), it breaks the U(1) gauge symmetry. 
The continuous gauge symmetry under which the
condensate (\ref{omega_omega_condensate_su2}) is invariant is therefore
\beq
{\rm SU}(4) \otimes {\rm SU}(2) \ .
\label{su4su2}
\eeq
This residual symmetry group has order 38 and rank 6.  For this and other
possible fermion condensation channels, we record these properties in Table
\ref{condensate_properties}. This table refers to the possible initial
condensation patterns at the highest condensation scale; subsequent evolution
further into the infrared is discussed below

A second possible condensation channel is 
\beq
\chi\chi: \quad 
([\bar 1]_6, [\bar 1]_2)_{-4} \times ([\bar 1]_6, [\bar 1]_2)_{-4} \to 
([\bar 2]_6, 1)_{-8} \ , 
\label{2x2_to_1_su2}
\eeq
where the shorthand name $\chi\chi$ reflects the associated condensate, 
$\epsilon^{\alpha\beta}\langle 
\chi_{i,\alpha,L}^T C \chi_{j,\beta,L}\rangle$. 
Since SU(2) has only pseudoreal representations, this channel has the form $2
\times 2 \to 1$ with respect to SU(2). The measure of attractiveness of this
channel due to the strongly coupled SU(2) gauge interaction is
\beq
\Delta C_2 = 2C_2([\bar 1]_2) = \frac{3}{2} \quad 
{\rm for} \ 2 \times 2 \to 1 \ {\rm in} \ {\rm SU}(2) \ . 
\label{Delta_C2_2x2_to_1_su2}
\eeq
From (\ref{alfcrit}), we find that the minimal critical coupling for
condensation in this channel is $\alpha_{cr} \simeq 4\pi/9 = 1.4$.  From the 
general structural analysis of fermion condensates given above, it follows
that, since the SU(2) tensor $\epsilon^{\alpha\beta}$ is antisymmetric, 
the condensate must be of the form $(a,a,s)$. (It cannot be of the form 
$(s,a,a)$ because with $N_f=1$, this would vanish identically.)  Hence, under
SU(6), it transforms as $[4]_6$, or equivalently, as $[\bar 2]_6$, as indicated
in Eq. (\ref{2x2_to_1_su2}).  Consequently, it is proportional to
\beq
\epsilon^{ijk\ell mn} \epsilon^{\alpha\beta} 
\langle \chi_{m,\alpha,L}^T C \chi_{n,\beta,L} \rangle \ , 
\label{chi_chi_condensate_su2}
\eeq
where $i,j,k,\ell,m,n$ are SU(6) indices and $\alpha,\beta$ are SU(2) indices.
This condensation channel thus preserves SU(2) while breaking U(1). As regards
SU(6), from a general group-theoretic analysis \cite{kim82,cgtakfb}, one
infers that the condensate (\ref{chi_chi_condensate_su2}) breaks this SU(6)
gauge symmetry to the subgroup ${\rm SU}(4) \otimes {\rm SU}(2)'$,
where we mark the ${\rm SU}(2)'$ with a prime to distinguish it from the SU(2)
in the original gauge group (\ref{gn6}). Hence, the full continuous gauge
symmetry under which the
condensate (\ref{chi_chi_condensate_su2}) is invariant is
\beq
[{\rm SU}(4) \otimes {\rm SU}(2)'] \otimes {\rm SU}(2) \ ,
\label{su4xsu2xsu2}
\eeq
where we insert the brackets to indicate the origin of the 
$[{\rm SU}(4) \otimes {\rm SU}(2)']$ group from the breaking of the original
SU(6) in (\ref{gn6}). This residual symmetry group has order 21 and rank 5. 

A third type of condensation channel is 
\beq
\chi\omega: \quad 
([\bar 1]_6, [\bar 1]_2)_{-4} \times (1,Adj)_6 \to ([\bar 1]_6,[1]_2)_2 \ , 
\label{2x3_to_2_su2}
\eeq
where the shorthand name $\chi\omega$ reflects the condensate 
\beq
\langle \chi_{i,\alpha,L}^T C \omega^{\alpha\beta}_{L} \rangle \ . 
\label{chi_omega_condensate_su2}
\eeq
With respect to SU(2), this channel is $2 \times 3 \to 2$. 
The measure of attractiveness for this channel due to SU(2) gauge interactions
is
\beq
\Delta C_2 = C_2(Adj) = 2 \quad 
{\rm for} \ 2 \times 3 \to 2 \ {\rm in} \ {\rm SU}(2) \ . 
\label{Delta_C2_2x3_to_2_su2}
\eeq
The corresponding estimate of the critical coupling from Eq. (\ref{alfcrit}) is
$\alpha_{cr}=\pi/3$.  Evidently, this channel is more attractive than the
$\chi\chi$ channel (\ref{2x2_to_1_su2}), but less attractive than the 
$\omega\omega$ channel (\ref{3x3_to_1_su2}). 

All of these three types of fermion condensation exhibit the phenomenon of a
strongly coupled gauge interaction producing condensate(s) that dynamically
break a more weakly coupled gauge interaction, namely U(1).  Furthermore, the
condensate in the $\chi\chi$ channel (\ref{2x2_to_1_su2}) dynamically breaks
not only the U(1) gauge symmetry, but also the more weakly coupled SU(6) gauge
symmetry. If a condensate were to form in the $\chi\omega$ channel
(\ref{2x3_to_2_su2}), it would self-break the strongly coupled SU(2) symmetry,
as well as breaking the weakly coupled SU(6) symmetry.  However, as will be
shown below, a condensate is not likely to form in the $\chi\omega$ 
channel. 

Since the $\omega\omega$ channel (\ref{3x3_to_1_su2}) is the MAC, one expects
that, as this theory evolves from the UV to the IR, at a scale that we denote
$\mu=\Lambda_1$ where the running coupling $\alpha_2(\mu)$ increases above the
critical value for condensation in this $\omega\omega$ channel, the condensate
(\ref{omega_omega_condensate_su2}) forms, breaking the U(1) gauge symmetry, but
leaving the SU(2) and SU(6) symmetries intact.  As the condensate 
$\langle {\vec \omega}^T_{L} C \, \cdot {\vec \omega}_{L}\rangle$ in 
Eq. (\ref{omega_omega_condensate_su2}) maintains the SU(2) symmetry, all of the
three components of the fermion ${\vec \omega}_L$ involved in this condensate
gain equal dynamical masses $\sim
\Lambda_1$ and are integrated out of the low-energy effective field theory that
describes the physics as the reference scale $\mu$ decreases below $\Lambda_1$.
The U(1) gauge field gains a mass $\sim g_3(\Lambda_1) \, \Lambda_1$. With
these fermion and vector boson fields integrated out, the one-loop and two-loop
coefficients in the SU(2) beta function in the low-energy effective theory have
the same sign, so as the reference momentum scale $\mu$ decreases below
$\Lambda_1$, the coupling $\alpha_2(\mu)$ continues to increase. Because the
${\vec \omega}_L$ fermions have been integrated out at the scale $\Lambda_1$,
they are no longer available to form a condensate in the $\chi\omega$ channel
(\ref{2x3_to_2_su2}) in the low-energy effective theory below $\Lambda_1$. 


\subsection{EFT Below $\Lambda_1$ and Condensation at 
Scale $\Lambda_2$ } 

In Ref. \cite{cgtsab} it was proved that if one
starts with a chiral gauge theory with gauge group $G$ that is free of gauge
and global anomalies, and it is broken dynamically to a theory with gauge group
$H \subseteq G$, with some set of fermions gaining dynamical masses and being
integrated out, then the low-energy theory with the gauge group $H$ is also
free of gauge and global anomalies. As a special case of this theorem, the
low-energy theory that is operative here as $\mu$ decreases below $\Lambda_1$
is also an anomaly-free theory. One easily checks that it is chiral.

As $\mu$ decreases below a lower scale that we denote as $\Lambda_2$,
$\alpha_2(\mu)$ increases past the critical value for the attractive $\chi\chi$
condensation channel (\ref{2x2_to_1_su2}), which is the MAC in this low-energy
effective theory, and the condensate (\ref{chi_chi_condensate_su2}) is expected
to form.  As noted above, this leaves SU(2) invariant and breaks SU(6) to ${\rm
  SU}(4) \otimes {\rm SU}(2)'$. By convention, one may label the SU(6) indices
$i,j$ of the fermions in the condensate (\ref{chi_chi_condensate_su2}) as $m=5$
and $n=6$. Then the fermions $\chi_{5\alpha,L}$ and $\chi_{6\beta,L}$
that are involved in this condensate gain dynamical
masses of order $\Lambda_2$ and are integrated out of the low-energy effective
theory applicable for $\mu < \Lambda_2$.  Furthermore, the gauge fields in
the coset space ${\rm SU}(6)/[{\rm SU}(4) \otimes {\rm SU}(2)]$ gain dynamical
masses of order $g_1(\Lambda) \Lambda_2$.


\subsection{EFT Below $\Lambda_2$ and Further Condensation}

By the same theorem as before, this
low-energy theory is anomaly-free and one can again check that it is
chiral. The low-energy effective theory below $\Lambda_2$ thus has a gauge
symmetry $[{\rm SU}(4) \otimes {\rm SU}(2)'] \otimes {\rm SU}(2)$, where the
${\rm SU}(2)'$ arises from the breaking of the SU(6) and the second SU(2) was
present in the original theory. The fermions that have gained masses and have
been integrated out are no longer dynamical.  The elements of the residual
SU(4) subgroup of SU(6) operate on the indices $1 \le i \le 4$, while the
elements of ${\rm SU}(2)'$ operate on the indices $i=5,6$.  Thus, the massless
fermions in this effective field theory below $\Lambda_2$ are as follows, where
we categorize them with a three-component vector, indicating the
representations with respect to the group (\ref{su4xsu2xsu2}) in the indicated
order:

\begin{enumerate}

\item
$\psi^{ij}_L$ with $1 \le i, \ j \le 4$, forming a (self-conjugate)
$([2]_4,1,1)$ representation
of the group ${\rm SU}(4) \otimes {\rm SU}(2)' \otimes {\rm SU}(2)$
in (\ref{su4xsu2xsu2}),

\item 
$\psi^{i5}_L$ and $\psi^{i6}_L$, forming a $([1]_4,[1]_{2'}1,1)$ representation
of (\ref{su4xsu2xsu2}), 

\item 
$\chi_{i,\alpha,L}$ with $1 \le i \le 4$, forming a 
$([\bar 1]_4,1,[\bar 1]_2)$ representation of (\ref{su4xsu2xsu2}).

\end{enumerate}

In this low-energy EFT below $\Lambda_2$, the MAC for SU(4)-induced 
condensate formations is $[2]_4 \times [2]_4 \to 1$ with the self-conjugate 
$\psi^{ij}_L$ transforming as $[2]_4$ of SU4), producing the condensate
\beq
\sum_{i,j,k,\ell = 1}^4 
\epsilon_{ijk\ell}\langle \psi^{ij \ T}_L C \psi^{k\ell}_L\rangle \ . 
\label{psi_psi_condensate_su4_last}
\eeq
This is a singlet under the SU(4) gauge symmetry and is obviously invariant
under the two other gauge symmetries, ${\rm SU}(2)' \otimes {\rm SU}(2)$, since
the fermions in (\ref{psi_psi_condensate_su4_last}) are singlets under these
groups.  Let us denote the scale at which this condensate forms as
$\Lambda_3$. The SU(4)-induced condensation producing this condensate
(\ref{psi_psi_condensate_su4_last}) has $\Delta C_2 = 5$. The $\psi^{ij}_L$
with $1 \le i,j \le 4$ involved in this condensate pick up dynamical masses of
order $\Lambda_3$ and are integrated out of the low-energy EFT that is
operative below $\Lambda_3$. The SU(4) gauge interaction can also produce the
condensate 
\beq
\sum_{i=1}^4 \langle \psi^{ir \ T}_L C \chi_{i,\alpha}\rangle \ , 
\label{psi_chi_condensate_last}
\eeq
where $r=5, \ 6$. This condensate is invariant under SU(4) and breaks 
${\rm SU}(2)' \otimes {\rm SU}(2)$ (since it involves the uncontracted 
${\rm SU}(2)'$ index $r$ and the uncontracted SU(2) index $\alpha$). For this
condensation, $\Delta C_2 = 15/4$. With these condensates, only the SU(4) gauge
symmetry remains, and all SU(4)-nonsinglet fermions have picked up dynamical
masses. This vectorial SU(4) theory confines and produces a spectrum of
SU(4)-singlet bound state hadrons. 


\section{$N=6$ Theory with SU(6) Gauge Interaction Dominant}
\label{su6_dominant_section}


\subsection{RG Evolution from UV} 

In this section we analyze the $N=6$ theory for the case in which the SU(6)
gauge interaction becomes strongly coupled and is dominant over the weakly
coupled SU(2) (and U(1)) gauge interactions. We denote this as the SU6D
case. The one-loop and two-loop terms in the beta function were given above in
Eqs. (\ref{b1_su6}) and (\ref{b2_su6}).  For both of the cases allowed by the
requirement of asymptotic freedom for the SU(6) and SU(2) gauge interactions,
namely $N_f=1$ and $N_f=2$, these coefficients have the same sign, so that the
two-loop beta function of this SU(6) theory has no IR zero. Hence, as the scale
$\mu$ decreases from $\mu_{_{UV}}$ to the IR, $\alpha_1(\mu)$ increases until
it eventually exceeds the range of values where it can be calculated
perturbatively.


\subsection{Highest-Scale Condensation Channels}

We examine the various possible fermion condensation channels produced by the
strongly coupled SU(6) gauge interaction.  The first is the $\psi\psi$ channel
\beq
\psi\psi: \quad ([2]_6,1)_2 \times ([2]_6,1)_2 \to ([4]_6,1)_4 \approx 
([\bar 2]_6,1)_4 \ , 
\label{a2_a2_to_a2bar_su6}
\eeq
with associated condensate
\beq
\epsilon_{ijk\ell mn} \langle \psi^{k\ell \ T}_{p,L} C  
\psi^{mn}_{p',L} \rangle \ . 
\label{psi_psi_condensate_su6}
\eeq
This condensate is automatically symmetrized in the flavor indices.  Since it
transforms as a $[\bar 2]_6$ representation of SU(6), it breaks SU(6) to ${\rm
  SU}(4) \otimes {\rm SU}(2)'$. Because the constituent fermion fields in
(\ref{psi_psi_condensate_su6}) are singlets under SU(2), this condensate is
obviously SU(2)-invariant. Finally, owing to the property that the condensate
(\ref{psi_psi_condensate_su6}) has nonzero U(1) charge, it also breaks
U(1). The residual subgroup of the original group (\ref{gn6}) that is left
invariant by the condensate (\ref{psi_psi_condensate_su6}) is thus $[{\rm
  SU}(4) \otimes {\rm SU}(2)'] \otimes {\rm SU}(2)$ (see Eq.
(\ref{su4xsu2xsu2})), as in the condensation process (\ref{2x2_to_1_su2}). The
condensation (\ref{a2_a2_to_a2bar_su6}) thus provides another example of an
induced, dynamical breaking of one gauge symmetry, namely U(1), by a different,
strongly coupled, gauge interaction in a direct-product chiral gauge theory.
The measure of attractiveness of this condensation channel involving the SU(6)
gauge interaction is
\beqs
\Delta C_2 &=& C_2([\bar 2]_6) = \frac{14}{3} \cr\cr
       && {\rm for} \ [2]_6 \times [2]_6 \to [\bar 2]_6 \ {\rm in} \ SU(6) . 
\label{Delta_C2_a2_a2_to a2bar_su6}
\eeqs
From the rough estimate for the minimal critical coupling strength to produce
this condensate, (\ref{alfcrit}), one has $\alpha_{cr} \simeq \pi/7 = 0.45$. 

A second possible condensation channel is 
\beq
\psi\chi: \quad 
([2]_6,1)_{2} \times ([\bar 1]_6,[\bar 1]_2)_{-4} \to ([1]_6,[\bar 1]_2)_{-2}
\label{a2_fbar_to_f_su6}
\eeq
with associated condensate
\beq
\langle \psi^{ij \ T}_{p,L} C \chi_{j, \beta,p',L} \rangle \ . 
\label{psi_chi_condensate_su6}
\eeq
This condensation breaks SU(6) to SU(5) and also breaks SU(2) and U(1), so that
the residual invariance group is SU(5), with order 24 and rank 4. The total
number of broken generators is thus 15 and the reduction in rank is by
3. Again, this illustrates the dynamical breaking of more weakly coupled gauge
symmetries by a strongly coupled gauge interaction in a direct-product gauge
theory.  The measure of attractiveness of this channel (\ref{a2_fbar_to_f_su6})
is
\beqs
\Delta C_2 &=& C_2([2]_6) = \frac{14}{3} \cr\cr
           &&{\rm for} \ 
[2]_6 \times [\bar 1]_6 \to [1]_6 \ {\rm in} \ SU(6) . 
\label{Delta_C2_a2_fbar_to_f_su6}
\eeqs
Evidently, this is the same as the attractiveness for the channel
(\ref{a2_a2_to_a2bar_su6}), so the critical coupling $\alpha_{cr}$ is also the
same as for that channel. This $\Delta C_2=14/3$ is also larger than the
$\Delta C_2$ for the third channel (to be discussed below), so that, as was
stated above in (\ref{su6macs}), for this $N=6$ theory, with SU(6) being the 
dominant gauge interaction, the $\psi\psi$ and $\psi\chi$ channels are the
MACs.  

A third condensation channel produced by the dominant SU(6) gauge interaction
is
\beq
\chi\chi: \quad
[\bar 1]_6 \times [\bar 1]_6 \to [\bar 2]_6 \approx [4]_6 \ \ {\rm in } \ \
{\rm SU}(6)
\label{fbar_fbar_a2bar_su6}
\eeq
with condensate 
\beq
\epsilon^{ijk\ell mn} \langle \chi_{m,\alpha,p,L}^T C
\chi_{n,\beta,p',L}\rangle \ . 
\label{chi_chi_condensate_su6}
\eeq
Although we use the same shorthand name, $\chi\chi$, for this channel as in
Eq. (\ref{2x2_to_1_su2}), it is understood that here it is the SU(6) gauge
interaction that is responsible for the formation of this condensate, rather
than the SU(2) gauge interaction in (\ref{2x2_to_1_su2}). 
The measure of attractiveness for this
condensation, as produced by the SU(6) gauge interaction, is
\beqs
\Delta C_2 &=& 2C_2([\bar 1]_6) -  C_2([2]_6) = \frac{7}{6} \cr\cr
&& {\rm for} \ [\bar 1]_6 \times [\bar 1]_6 \to [\bar 2]_6 \ 
{\rm in} \ {\rm SU}(6) 
\ . 
\label{Delta_C2_fbar_fbar_to_a2_su6}
\eeqs
This $\Delta C_2$ is a factor of 4 smaller than the common value $\Delta
C_2=14/3$ for the condensation channels (\ref{a2_a2_to_a2bar_su6}) and
(\ref{a2_fbar_to_f_su6}) and hence is predicted not to occur in this
SU(6)-dominant case.  We proceed to discuss in greater detail the two different
patterns of UV to IR evolution for the most attractive condensation channels in
this SU(6)-dominant case. 


\subsection{$\psi\psi$ Condensation Channel } 
\label{psi_psi_su6_section} 

Here we consider the $\psi\psi$ condensation channel
(\ref{a2_a2_to_a2bar_su6}), i.e., $([2]_6,1)_2 \times ([2]_6,1)_2 \to ([\bar
2]_6,1)_4$. We denote the scale at which the condensate
(\ref{psi_psi_condensate_su6}) forms as $\Lambda_1$. (To avoid cumbersome
notation, we use the same symbol for this highest-level condensation as we did
in the subsection dealing with the case where the SU(2) gauge interaction is
dominant, but it is understood implicitly that this scale has generically
different values for these different cases.) Without loss of generality, one
may choose the SU(6) group indices of the $\psi$ fermions involved in the
condensate (\ref{psi_psi_condensate_su6}) to be $k, \ \ell, \ m, \ n \in
\{1,2,3,4 \}$ and the uncontracted indices in (\ref{psi_psi_condensate_su6}) to
be $i,j \in \{5,6\}$. The $\psi$ fermions involved in the condensate
(\ref{psi_psi_condensate_su6}) gain dynamical masses of order $\Lambda_1$. The
gauge bosons in the coset ${\rm SU}(6)/[{\rm SU}(4) \otimes {\rm SU}(2)']$ pick
up dynamical masses of order $g_1(\Lambda_1) \, \Lambda_1$, and the U(1) gauge
boson picks up a dynamical mass $\simeq g_3(\Lambda_1) \, \Lambda_1$.  These
massive fermion and vector boson fields are integrated out of the low-energy
effective field theory that describes the physics as the reference scale $\mu$
decreases below $\Lambda_1$.  The resultant low-energy effective theory
contains the following massless fermions: (1) ${\rm SU}(4) \otimes {\rm
  SU}(2)'$-nonsinglets $\psi^{ia}_{p,L}$ with $1 \le i \le 4$, $a \in \{5,6\}$,
and $1 \le p \le N_f$, which are singlets under SU(2); (2) ${\rm SU}(4) \otimes
{\rm SU}(2)$-nonsinglets $\chi_{i \alpha,p,L}$ with $1 \le i \le 4$,
$\alpha=1,2$, and $1 \le p \le N_f$, which are singlets under ${\rm SU}(2)'$;
and (3) ${\rm SU}(2)' \otimes {\rm SU}(2)$-nonsinglets $\chi_{i,\alpha,p,L}$
with $i=5,6$, which are singlets under SU(4).  There are also the massless
fermions $\psi^{56}_{p,L}$ with $1 \le p \le N_f$, which are singlets under all
three factor groups in (\ref{su4xsu2xsu2}).  The fermions (1) transform as
$2N_f$ fundamental representations $F=[1]_4$ of SU(4), while the fermions (2)
transform as $2N_f$ conjugate fundamental representations $\bar F=[\bar 1]_4$
of SU(4), so that the SU(4) gauge symmetry is vectorial.  Combining this
property with the fact that the ${\rm SU}(2)'$ and SU(2) groups have only real
representations, it follows that this low-energy theory is vectorial. The
action of an element $U \in {\rm SU}(4)$ is
\beqs
\psi^{ia}_{p,L} &=& U^i_j \psi^{ja}_{p,L} \cr\cr
\chi_{ia,p,L} &=& (U^\dagger)^j_i \chi_{ja,p,L} \ , 
\label{su4psichi}
\eeqs
with fixed $a=5,6$ and $1 \le p \le N_f$. The elements of ${\rm SU}(2)'$
operate on the indices $a=5,6$ of the fermions (1) and (3). (The operation of
the elements of SU(2) on the $\alpha,\beta$ indices has already been
discussed.)  The couplings of the SU(4) and ${\rm SU}(2)'$ gauge interactions
start out equal at $\mu=\Lambda_1$, as descendents of the gauge coupling
$\alpha_1$ of the UV gauge coupling for the SU(6) gauge interaction.

As the theory evolves further into the IR, several possible patterns of gauge
symmetry breaking are possible.  The SU(4) gauge interaction can produce a 
condensate in the $[1]_4 \times [\bar 1]_4 \to 1$, i.e., $F \times \bar F \to
1$ channel: 
\beq
\langle \sum_{i=1}^4 \psi^{ia \ T}_{p,L} C \chi_{i,\alpha,p',L} \rangle \ ,
\label{psi_chi_su4}
\eeq
where, as indicated, the sum on $i$ is over the active SU(4) gauge indices,
while the other indices take on the values $a=5, \ 6$, $\alpha=1, \ 2$, and $1
\le p,p' \le N_f$. The measure of attractive of this condensation, as produced
by the SU(4) gauge interaction, is $\Delta C_2 = 2C_2([1]_4) = 15/4 = 3.75$.
This condensate preserves the SU(4) gauge symmetry
and breaks the ${\rm SU}(2)'$ gauge symmetry 
operating on the indices $a=5,6$ and the SU(2) gauge symmetry operating
on the indices $\alpha=1,2$. 

In contrast, the ${\rm SU}(2)'$ gauge interaction could produce the condensate
\beq
\sum_{a,b=5}^6 \epsilon_{ab} 
\langle \psi^{ia \ T}_{p,L} C \psi^{jb}_{p',L}\rangle \ . 
\label{psipsi_su4b}
\eeq
The measure of
attractiveness for this condensation, as produced by the 
${\rm SU}(2)'$ interaction, is $\Delta C_2 = 3/2$. 
Since the fermions involved in this condensate are SU(2)-singlets, it obviously
preserves SU(2).  With the contraction on the ${\rm SU}(2)'$ indices $a,b \in
\{5,6\}$, it also preserves ${\rm SU}(2)'$.  If $N_f=1$, then the condensate is
automatically symmetric in the single flavor index, so it has the form 
$(a,a,s)$ in the notation of Eq. (\ref{sss}) and hence transforms like the
$[2]_4$ representation of SU(4).  This breaks SU(4) to 
$SU(2)'' \otimes SU(2)'''$, where we use repeated primes to indicate that these
SU(2) subgroups of SU(4) are distinct from both the original UV SU(2) symmetry
and the ${\rm SU}(2)'$ symmetry. If $N_f=2$, then there are two possibilities;
$(a,a,s)$ if one constructs a linear combination that is symmetrized in 
flavor indices, and $(a,s,a)$, if one antisymmetrizes over flavor indices. 
For each of these possibilities, one can track the evolution further into the
IR using the same methods as above. 


\subsection{$\psi \chi$ Condensation Channel } 
\label{a2_fbar_to_f_su6_section} 

Here we consider the $\psi\chi$ condensation channel (\ref{a2_fbar_to_f_su6}),
i.e., $([2]_6,1)_2 \times ([\bar 1]_6,[\bar 1]_2)_{-4} \to ([1]_6,[\bar
1]_2)_{-2}$, with the associated condensate 
$\langle \psi^{ij \ T}_{p,L} C \chi_{j,\beta,p',L} \rangle$ in Eq. 
(\ref{psi_chi_condensate_su6}). By convention, we may choose the SU(6) index 
$i=6$ and the SU(2) index $\beta=2$ in this condensate. Then the fermions 
involved in the condensate, namely
$\psi^{6j}_{p,L}$ and $\chi_{j,2,p',L}$ with $1 \le j \le 5$ gain dynamical
masses of order $\Lambda_1$ and are integrated out of the
low-energy effective theory applicable for $\mu < \Lambda_1$.
The 11 SU(6) gauge bosons in the coset ${\rm SU}(6)/{\rm SU}(5)$ gain dynamical
masses of order $g_1(\Lambda_1) \Lambda_1$, while
the SU(2) and U(1) gauge bosons gain masses of order 
$g_i(\Lambda_1) \Lambda_1$ with $i=2,3$,
respectively. These fields are integrated out of the low-energy effective
theory applicable for $\mu < \Lambda_1$. 

For this channel, the low-energy effective theory that describes
the physics as $\mu$ decreases below $\Lambda_1$ has an SU(5) gauge symmetry
with (massless) SU(5)-nonsinglet fermions $\psi^{ij}_{p,L}$ and 
$\chi_{i,1,p',L}$, where $1 \le i,j \le 5$ and $1 \le p,p' \le N_f$.  
In addition, there are massless SU(5)-singlet
fermions $\chi_{6,\beta,p',L}$ and $\omega^{\alpha\beta}_{p,L}$ with 
$1 \le \alpha, \beta \le 2$ and $1 \le p,p' \le N_f$ remaining from
the UV theory.  

In this low-energy theory, the SU(5) gauge coupling inherited
from the SU(6) UV theory continues to increase as $\mu$ decreases below
$\Lambda_1$, and is expected to trigger a further fermion condensation
\beq
[2]_5 \times [2]_5 \to [\bar 1]_5 
\label{a2_a2_to_fbar_su5}
\eeq
with $\Delta C_2 = 24/5$ and associated condensate 
\beq
\sum_{i,j,k,\ell,m=1}^5 \, 
\epsilon_{ijk\ell m} \langle \psi^{jk \ T}_{p,L} C \psi^{\ell m}_{p',L}\rangle
\ , 
\label{psi_psi_condensate_su5}
\eeq
where the indices $i, \ j, \ k, \ell, m$ are SU(5) group indices.  By
convention, we may choose the uncontracted SU(5) group index in
(\ref{psi_psi_condensate_su5}) to be $i=5$.  This condensate breaks SU(5) to
SU(4).  The fermions $\psi^{jk}_{p,L}$ with $j,k \in \{1,2,3,4\}$ and $1 \le p
\le N_f$ gain dynamical masses of order $\Lambda_2$. The 9 gauge bosons in the
coset SU(5)/SU(4) gain dynamical masses of order $g_1(\Lambda_2)\Lambda_2$.
All of these fields are integrated out of the low-energy effective theory that
describes the physics at scales $\mu < \Lambda_2$.

The low-energy theory that is operative for $\mu < \Lambda_2$ has a gauge group
SU(4) and (massless) SU(4)-nonsinglet fermion content consisting of
$\psi^{ij}_{p,L}$ with $1 \le i,j \le 4$ and $1 \le p \le N_f$. However, this
representation, $[2]_4$, in SU(4) is self-conjugate, i.e., $[2]_4 \approx [\bar
2]_4$, so this theory is vectorial. The two-loop beta function for this theory
has no IR zero and as $\mu$ continues to decrease, the SU(4) coupling inherited
from the SU(5) theory continues to increase. Because of the vectorial nature of
this descendent SU(4) theory, the condensate that forms is in the channel
$[2]_4 \times [2]_4 \to 1$, with condensate
\beq
\sum_{i,j,k,\ell=1}^4 \, 
\epsilon_{ijk\ell} \langle \psi^{ij \ T}_{p,L} C \psi^{k\ell}_{p',L}\rangle 
\ , 
\label{psi_psi_condensate_su4}
\eeq
with $\Delta C_2=5$, where here, $i, \ j, \ k, \ \ell$ are SU(4) group
indices. This condensate preserves the SU(4) gauge symmetry, while breaking
global chiral symmetries spontaneously. The fermions involved in this
condensate pick up dynamical masses of order the condensation scale.  This
theory confines and produces a spectrum of SU(4)-singlet bound state hadrons. 


\section{$N=6$ Theory with SU(6) and SU(2) Gauge Interactions Comparable in 
Strength}
\label{n6_both_section}


\subsection{General Discussion} 

In this section we consider the situation in which both the SU(6)
and SU(2) gauge interactions are of comparable strength and hence must be
treated together (with the U(1) gauge interaction still being weak). 
In this case, one cannot neglect the mixing terms at the two-loop and
higher-loop level in the beta functions $\beta_{\alpha_i}$, Eq. (\ref{beta}),
so the calculation the evolution of the gauge couplings down from the 
initial reference point $\mu=\mu_{_{UV}}$ in the UV is more complicated. 
For our present purposes, it will suffice to consider a case in which 
$\alpha_1(\mu) \simeq \alpha_2(\mu) \simeq O(1)$ at a lower scale $\mu$. Since
the SU(2) interaction by itself would evolve to a relatively weakly coupled
IRFP if $N_f=2$, expected to be in the non-Abelian Coulomb phase, we will
assume $N_f=1$ here, to guarantee that not just the SU(6) interaction, but also
the SU(2) interaction become strongly coupled in the infrared. 


\subsection{Analysis of Possible Condensation Channels}

\subsubsection{Condensation(s) Involving SU(6)-Nonsinglet Fermions}

We have shown above that the most attractive condensation channels are
different in the simple situations where either the SU(6) or the SU(2) gauge
interactions are dominant.  Specifically, in the SU(2)-dominant case, the MAC
is the $\omega\omega$ channel, with $\Delta C_2=4$, while in the SU(6)-dominant
case, the MACs are the $\psi\psi$ and $\psi\chi$ channels, with the same
measure of attractiveness, $\Delta C_2=14/3 = 4.7$. One would thus expect that
as the reference scale decreases, the first condensate(s) to form would be in
the $\psi\psi$ and/or $\psi\chi$ channels, as produced by the SU(6) gauge
interaction.  Since the $\psi\psi$ channels involves SU(2)-singlet fermions, it
would not be affected by the fact that the SU(2) gauge interaction is also
strongly coupled. The other SU(6) MAC, namely the $\psi\chi$ channel involves
the SU(2)-singlet fermion $\psi$ and the SU(2)-nonsinglet fermion $\chi$, so
the binding is only caused by the SU(6) interaction. 
Since the $\psi\chi$ condensation leaves the residual gauge
symmetry group SU(5), of order 24, while the $\chi\chi$ condensation would
leave the residual gauge symmetry (\ref{su4xsu2xsu2}), of order 21, a vacuum
alignment argument suggests that the $\psi\chi$ condensation channel is 
preferred over the $\chi\chi$ channel.  Thus, the $\psi\chi$ condensate 
(\ref{psi_chi_condensate_su6}) is expected to form at a scale that we will
denote as $\Lambda_1$, self-breaking SU(6) to SU(5) and also producing 
induced dynamical breaking of U(1).  The 11 gauge bosons in the coset 
${\rm SU}(6)/{\rm SU}(5)$ gain dynamical masses of order 
$g_1(\Lambda_1) \, \Lambda_1$, and the U(1) gauge boson gains a mass of order 
$g_3(\Lambda_1) \, \Lambda_1$. 


\subsubsection{EFT Below $\Lambda_1$}

Next, one would expect that condensation would occur in the
$\omega\omega$ channel, as produced by the strong SU(2) gauge
interaction. Owing to the fact that the value of $\Delta C_2$ for this
condensation is equal to 4, slightly less than the value of 4.7 for the
$\psi\psi$ condensation, one expects that this occurs at a slightly lower
scale. Because this second condensation would give dynamical masses to
the $\omega$ fermions, which would thus be integrated out of the low-energy
theory applicable below this condensation scale, it would preclude the
formation of an SU(2)-induced condensate in the $\chi\omega$ channel.  

There remains the $\chi\chi$ condensation channel.  Although the value of
$\Delta C_2$ for this condensation, as produced by the SU(6) interaction, is
7/6, which is a factor of 4 smaller than the value of 14/3 for the MACs, and
although the value of $\Delta C_2$ for this condensation, as produced by the
SU(2) interaction, is 3/2, considerably smaller than the value $\Delta C_2 = 4$
for the SU(2)-induced MAC channel, $\omega\omega$, the $\chi\chi$ channel has
the special property that it involves both the SU(2) and SU(6) gauge
interactions, in contrast to all of the other possible condensation channels
($\psi\psi$, $\psi\chi$, $\omega\omega$, and $\chi\omega$), each of which only
involves one of these two non-Abelian gauge interactions.  
If $\alpha_1(\mu)=\alpha_2(\mu)$ and one were simply to 
add the two terms $(7/6)\alpha_1(\mu)+(3/2)\alpha_2(\mu) =
(8/3)\alpha_1(\mu)$, the effective $\Delta C_2$ would be 8/3 = 3.7, which is 
still less than values for the MACs for both the SU(6)-induced condensates 
and the SU(2)-induced condensates. 


\section{Related Constructions}
\label{related_constructions_section}

At the beginning of this paper we remarked on how the theory (\ref{ggen}) with
(\ref{fermions}) successfully combines two different (anomaly-free) chiral
gauge theories, SU($N$) with $N_f$ copies of (\ref{a2fb}), and SU($M$) with
$N_f$ copies of (\ref{s2fb}), where $M=N-4$.  A natural question concerns
related constructions of direct-product chiral gauge theories with fermions in
higher-rank tensor representations of the factor groups. The next step up in
complexity involves rank-3 antisymmetric and symmetric tensor representations
for the fermions. Two theories with these rank-3 representations use a gauge
group of the form
\beq
{\rm SU}(N) \otimes {\rm SU}(M) \otimes {\rm U}(1) \ , 
\label{sunxsumxu1}
\eeq
where now $M$ can take on two different values as a function of $N$, namely 
$M=N-3$ or $N=N-6$. In both cases, the fermion content consists of 
$N_f$ copies of the set \cite{barr2015,anomcal} 
\beq
([3]_N,1)_{q_{30}} + ([\bar 2]_N,(\bar 1)_M)_{q_{21}} + ([1]_N,(2)_M)_{q_{12}}
+ (1,(\bar 3)_M)_{q_{03}}
\label{fermions_k3}
\eeq
with 
\beqs
M=N-3 \ & \Longrightarrow & (q_{30},q_{21},q_{12},q_{03}) = \cr\cr
&=& \Big ( -(N-3),(N-2),-(N-1),N \Big ) \cr\cr
&&
\label{qu1_nm3}
\eeqs
and
\beqs
M=N-6 \ & \Longrightarrow & (q_{30},q_{21},q_{12},q_{03})= \cr\cr
&=& \Big (-(N-6),(N-4),-(N-2),N \Big ) \ . \cr\cr
&&
\label{qu1_nm6}
\eeqs
Owing to the presence of the factor group SU($M$) in (\ref{sunxsumxu1}), 
the lowest nondegenerate cases are $N=5$ if $M=N-3$ and $N=8$ if
$M=N-6$. 

As before, there are equivalent theories. One has all of the 
representations of the (left-handed chiral) fermions conjugated.  The second
has the SU($M$) representations conjugated relative to the SU($N$)
representations, i.e., it has a fermion content comprised of $N_f$ copies of 
the set 
\beq
([3]_N,1)_{q_{30}} + ([\bar 2]_N,(1)_M)_{q_{21}} + ([1]_N,(\bar 2)_M)_{q_{12}}
+ (1,(3)_M)_{q_{03}} \ . 
\label{fermions_k3_twisted}
\eeq
Since these are equivalent to the theory with gauge group 
(\ref{sunxsumxu1}) and fermions (\ref{fermions_k3}), with the indicated
U(1) charges for $M=N-3$ and $M=N-6$, it suffices to discuss only the latter
theories.

However, none of these theories satisfies the requisite condition for our
analysis, that both the SU($N$) and SU($M$) gauge interactions are
asymptotically free (AF).  The reason for this is as follows.  In the theory
(\ref{ggen}) with (\ref{fermions}), the one-loop term in the SU($N$) and
SU($N-4$) beta functions involves the trace invariants for the fundamental and
symmetric or antisymmetric rank-2 representations. While $T([2]_N)=(N-2)/2$ and
$T((2)_N)$ are linear functions of $N$ and hence enter the one-loop
coefficients in the beta functions with the same polynomial degree as the pure
gauge contribution, $T([3]_N)$ and $T((3)_M)$ are quadratic functions of $N$
and $M$, respectively, namely $T([3]_N=(N-3)(N-4)/4)$ and
$T((3)_M)=(M+3)(M+4)/4$.  The most stringent restriction arises from the
constraint that the SU($M$) beta function be negative.  The one-loop
coefficient in this beta function is
\beqs
b^{({\rm SU}(M))}_{1\ell;11} &=& \frac{1}{3}\bigg [ 11M - N_f \bigg \{
\frac{N(N-1)}{2} + N(M+2) + \cr\cr
&+& \frac{(M+2)(M+3)}{2} \bigg \} \bigg ] \ . 
\label{b1_sum_k3}
\eeqs
For the theory with $M=N-3$, this is
\beq
b^{({\rm SU}(N-3))}_{1\ell;11} = \frac{1}{3}\Big [11(N-3)-2N_fN(N-1)\Big ] \ .
\label{b1_sum_k3a}
\eeq
This one-loop coefficient is negative if $N_f > N_{f,b1z,k3a}$, where 
\beq
N_{f,b1z,k3a} = \frac{11(N-3)}{2N(N-1)} \ . 
\label{nfb1z_sumk3a}
\eeq
We find that $N_{f,b1z,k3a} < 1$ for all $N$ in the relevant range $N \ge 5$. 
Hence, the AF constraint does not allow any nonzero value of $N_f$. 
Similarly, for the theory with $M=N-6$,
\beq
b^{({\rm SU}(N-6))}_{1\ell;11} = \frac{1}{3}\Big [11(N-6)-2N_f(N^2-4N+3) 
\Big ] \ . 
\label{b1_sum_k3b}
\eeq
This one-loop coefficient is negative if $N_f > N_{f,b1z,k3b}$, where
\beq
N_{f,b1z,k3b} = \frac{11(N-6)}{2N(N^2-4N+3)} \ . 
\label{nfb1z_sumk3b}
\eeq
The value of $N_{f,b1z,k3b}$ is less than 1 for all $N$ in the relevant range,
$N \ge 8$. Therefore, the AF constraint does not allow any nonzero value of
$N_f$.  We recall that $N_f$ must be nonzero in order for the theory to be a
chiral gauge theory, since if $N_f=0$, then the theory degenerates into
decoupled purely gluonic sectors. Thus, in neither of these theories with
rank-3 fermion representations and $M=N-3$ or $M=N-6$ is the SU($M$) gauge
interaction asymptotically free.  Similar comments apply to ${\rm SU}(N)
\otimes {\rm SU}(M) \otimes {\rm U}(1)$ theories with fermions in sets of
representations containing antisymmetric and symmetric rank-$k$ tensor
representations of the non-Abelian gauge groups with $k \ge 4$. As was
discussed above, the requirement of asymptotic freedom of both of the
non-Abelian gauge interactions was imposed because of (i) the purpose of
studying the strong-coupling behavior of one or both of these interactions as
the theory evolves from the UV to the IR and (ii) the necessity to be able to
carry out a self-consistent perturbative calculation of the beta functions for
these interactions at a reference scale, $\mu_{_{UV}}$.


\section{Conclusions}
\label{conclusions_section}

In nature, the ${\rm SU}(2)_L \otimes {\rm U}(1)_Y$ electroweak symmetry is
broken not only by the vacuum expectation value of the Higgs field, but also
dynamically, by the $\langle \bar q q \rangle$ quark condensates produced by
the color SU(3)$_c$ gauge interaction. Moreover, sequential self-breakings
of strongly coupled chiral gauge symmetries have also been used in models of
dynamical generation of fermion masses. In this paper we have investigated a
chiral gauge theory that serves as a theoretical laboratory that exhibits both
induced breaking of a weakly coupled gauge symmetry via condensates formed
by a different, strongly coupled gauge interaction, and also self-breaking of
strongly coupled chiral gauge symmetries. We have studied an asymptotically
free chiral gauge theory with the direct-product gauge group ${\rm SU}(N)
\otimes {\rm SU}(N-4) \otimes {\rm U}(1)$ and chiral fermion content consisting
of $N_f$ flavors of fermions transforming according to the representations
$([2]_N,1)_{N-4} + ([\bar 1]_N,[\bar 1]_{N-4})_{-(N-2)} + (1,(2)_{N-4})_N$.
One of the reasons for interest in this theory is that it may be viewed as a
combination of two separate (anomaly-free) chiral gauge theories, namely (i) an
SU($N$) theory with fermion content consisting of $N_f$ flavors of fermions in
the $[2]_N$ and $N-4$ copies of $[\bar 1]_N$, and (ii) an SU($M$) theory with
fermions consisting of $N_f$ flavors of fermions in the $(2)_M$ and $M+4$
copies of $[\bar 1]_M$, with $M=N-4$, which also incorporates a U(1) gauge
symmetry. We have analyzed the UV to IR evolution of this theory and have
investigated patterns of possible bilinear condensate formation.  A detailed
discussion of the lowest nondegenerate case, $N=6$ was given.  This analysis
involved a sequential construction and analysis of low-energy effective field
theories that describe the physics as the theory evolves through various 
condensation scales and certain fermions and gauge bosons pick up dynamically
generated masses. Our findings provide new insights into the phenomenon of
induced breaking of a weakly coupled gauge symmetry by a different, strongly
coupled gauge interaction, and self-breaking of a strongly coupled chiral gauge
symmetry.


\begin{acknowledgments}

This research was supported in part by the Danish National
Research Foundation grant DNRF90 to CP$^3$-Origins at SDU (T.A.R.) and
by the U.S. NSF Grant NSF-PHY-16-1620628 (R.S.)

\end{acknowledgments}



\newpage

\begin{sidewaystable}
  \caption{\footnotesize{Properties of possible initial (highest-scale) 
     bilinear fermion 
    condensates in the UV theory (\ref{ggen}) with (\ref{fermions}) for 
    $N \ge 7$. 
      The shorthand name of the condensation channel is listed in the first
      column, and the corresponding condensate is displayed in 
      the the second column. The third and fourth columns list 
      the values of $\Delta C_2$ with respect to the SU($N$) and SU($N-4$) 
      gauge
      interactions. The entries in the fifth, sixth, and seventh columns
      indicate whether a given condensate is invariant (inv.) under the 
      SU($N$), SU($N-4$), and U(1) gauge symmetries, respectively, 
      or breaks (bk.) 
      one or more of these symmetries. The entry in the eighth column 
      gives the representation $(R_1,R_2)_q$ of the condensate under the group 
      (\ref{ggen}), following the notation of Eq. (\ref{rrq}). The ninth 
      column lists the continuous gauge 
      symmetry group under which a given condensate is invariant. 
      The tensors $\epsilon_{...k\ell mn}$ and $\epsilon^{...mn}$ are
      antisymmetric SU($N$) tensors, while $\epsilon^{...\alpha\beta}$ is an
      antisymmetric SU($N-4$) tensor. These 
      results are for the case $N_f=1$ and for $N_f \ge 2$ with condensates
      symmetrized over the flavor indices, which are suppressed in the
      notation. The $\psi\chi$ channel is the MAC for the SU($N$)-dominant
      case, while the $\chi\omega$ channel is the MAC for the
      SU($N-4$)-dominant case in this $N \ge 7$ range. 
      See text for further discussion.}}
  \footnotesize
\begin{center}
\begin{tabular}{|c|c|c|c|c|c|c|c|c|} \hline\hline
name & condensate & $(\Delta C_2)_{{\rm SU}(N)}$ & 
$(\Delta C_2)_{{\rm SU}(N-4)}$ & 
${\rm SU}(N)$ & ${\rm SU}(N-4)$ & U(1) & $(R_1,R_2)_q$ & $H_{inv}$ \\
\hline
$\psi\chi$ & 
$\langle \psi^{ij \ T}_L C \chi_{j,\beta,L} \rangle$ & 
$\frac{(N-2)(N+1)}{N}$ & 0 & bk.  & bk. & bk. & 
$([1]_N,[\bar 1]_{N-4})_{-2}$ & ${\rm SU}(N-1) \otimes {\rm SU}(N-5)$ \\

$\chi\omega$ & 
$\langle \chi_{i, \alpha,L}^T C \omega^{\alpha \beta}_L\rangle$ &
0 & $\frac{(N-2)(N-5)}{N-4}$ & bk.  & bk.  & bk. & 
$([\bar 1]_N,[1]_{N-4})_2$ & ${\rm SU}(N-1) \otimes {\rm SU}(N-5)$ \\

$\psi\psi$ & 
$\epsilon_{...k\ell mn} \langle \psi^{k\ell \ T}_L C \psi^{mn}_L \rangle$ & 
$\frac{4(N+1)}{N}$ & 0 & bk.  & inv. & bk. & $([4]_N,1)_{2(N-4)}$ & 
$[{\rm SU}(N-4) \otimes {\rm SU}(4)] \otimes {\rm SU}(N-4)$ \\

$\chi\chi$ & 
$\epsilon^{...mn} \epsilon^{...\alpha\beta} \langle 
\chi_{m,\alpha,L}^T C \chi_{n,\beta,L} \rangle$ & 
$\frac{N+1}{N}$ & $\frac{N-3}{N-4}$ & bk.  & bk. & bk. &  
$([\bar 2]_N,[\bar 2]_{N-4})_{-2(N-2)}$ & 
$[{\rm SU}(N-2) \otimes {\rm SU}(2)'] \otimes$ \\ 
& & & & & & & & $\otimes [{\rm SU}(N-6) \otimes {\rm SU}(2)'']$ \\
\hline\hline
\end{tabular}
\end{center}
\label{condensate_properties_sun}
\end{sidewaystable}
%


\begin{sidewaystable}
  \caption{\footnotesize{Properties of possible initial bilinear fermion 
    condensates in the UV theory (\ref{gn6}) with (\ref{fermions_n6}). 
    The shorthand name of the
    condensation channel and the condensate in this channel are displayed in
      the the first and second columns. The third and fourth columns list 
      the values of $\Delta C_2$ with respect to the SU(6) and SU(2) gauge
      interactions. The entries in the fifth, sixth, and seventh columns
      indicate whether a given condensate is invariant (inv.) under the SU(6),
      SU(2), and U(1) gauge symmetries, respectively, 
      or breaks (bk.) one or more of these
      symmetries. The entry in the eighth column gives the representation 
      $(R_1,R_2)_q$ of the condensate under the group 
      (\ref{gn6}), following the notation of Eq. (\ref{rrq}). The ninth
      column lists the continuous gauge 
      symmetry group under which a given condensate is invariant. These 
      results are for the case $N_f=1$ and for $N_f=2$ with condensates
      symmetrized over the flavor indices, which are suppressed in the
      notation. The $\psi\psi$ and $\psi\chi$ channels are the MACs for the
      SU(6)-dominant case, while the $\omega\omega$ is the MAC for the
      SU(2)-dominant case. See text for further discussion.}}
\begin{center}
\begin{tabular}{|c|c|c|c|c|c|c|c|c|} \hline\hline
name & condensate & $(\Delta C_2)_{{\rm SU}(6)}$ & 
                    $(\Delta C_2)_{{\rm SU}(2)}$ & 
SU(6) & SU(2) & U(1) & $(R_1,R_2)_q$ & $H_{inv}$ \\ 
\hline
$\omega\omega$& 
$\langle {\vec \omega}_L^T C \cdot {\vec \omega}_L \rangle$ &  
0 & 4 & inv. & inv. & bk. & $(1,1)_{12}$ & ${\rm SU}(4)\otimes {\rm SU}(2)$ \\

$\psi\chi$ & 
$\langle \psi^{ij \ T}_L C \chi_{j,\beta,L} \rangle$ & 
$\frac{14}{3}$ & 0 & bk.  & bk. & bk. & $([1]_6,[\bar 1]_2)_{-2}$ & SU(5) \\

$\chi\omega$ & 
$\langle \chi_{i, \alpha,L}^T C \omega^{\alpha \beta}_L \rangle$ &
0 & 2 & bk.  & bk.  & bk. & $([\bar 1]_6,[1]_2)_2$ & SU(5) \\

$\psi\psi$ & 
$\epsilon_{ijk\ell mn}\langle\psi^{k\ell\ T}_L C \psi^{mn}_L \rangle$ & 
$\frac{14}{3}$ & 0 & bk. & inv. & bk. & $([\bar 2]_6,1)_4$ & 
$[{\rm SU}(4)\otimes {\rm SU}(2)'] \otimes {\rm SU}(2)$ \\

$\chi\chi$ & 
$\epsilon^{ijk\ell mn} \epsilon^{\alpha\beta} \langle 
\chi_{m,\alpha,L}^T C \chi_{n,\beta,L} \rangle$ & 
$\frac{7}{6}$ & $\frac{3}{2}$ & bk.  & inv. & bk. & $([\bar 2]_6,1)_{-8}$ &  
$[{\rm SU}(4)\otimes {\rm SU}(2)'] \otimes {\rm SU}(2)$ \\

\hline\hline
\end{tabular}
\end{center}
\label{condensate_properties}
\end{sidewaystable}
%



\begin{thebibliography}{99}
\bibitem{refs}
%
  See, e.g., Ref. \cite{cgt} 
  for references to the literature on chiral gauge theories with a single gauge
  group. Here we only deal with theories with direct-product chiral gauge
  groups.
\bibitem{casimir}
%
  We recall the definitions of relevant group invariants.  Let $T_a$ denote the
  generators of a Lie algebra of $G$, and let $d(R)$ denote the dimension of a
  representation $R$ of $G$. Then the quadratic Casimir invariant $C_2(R)$ is
  defined by $\sum_a T^a_RT^a_R = C_2(R) I$, where $I$ is the $d_R \times d_R$
  identity matrix. The trace invariant $T(R)$ is defined by ${\rm Tr}_R(T^a_R
  T^b_R) = T(R)\delta^{ab}$. We use standard normalizations for SU($N$) so
  that, e.g., for the fundamental representation $[1]_N$, $T([1]_N)=1/2$ and
  $C_2([1]_N)=(N^2-1)/(2N)$.
\bibitem{vacalign}
%
  In passing, we recall that a vacuum alignment argument explains a basic
  property of the dynamical breaking of $G_{EW}$ to U(1)$_{em}$ by quark
  condensates.  At the level of SU(3)$_c$ by itself, the quark condensates
  $\langle \bar u d \rangle$ and $\langle \bar u s \rangle$ are of the same
  $\bar 3 \times 3 \to 1$ form as the $\langle \bar u u \rangle$, $\langle \bar
  d d \rangle$, and $\langle \bar s s \rangle$ condensates and have the the
  same attractiveness measure, $\Delta C_2=8/3$.  However, if $\langle \bar u d
  \rangle$ or $\langle \bar u s \rangle$ were nonzero, this would violate
  U(1)$_{em}$. The vacuum alignment precludes the 
  $\langle \bar u d \rangle$ and $\langle \bar u s \rangle$ condensates and
  thus preserves U(1)$_{em}$ as the maximal residual gauge symmetry.
\end{thebibliography}
\end{document}